\documentclass{article}

\usepackage[round]{natbib}
\usepackage[T1]{fontenc}
\usepackage[utf8]{inputenc}

\usepackage[normalem]{ulem}

\usepackage{graphicx}

\usepackage{amsmath,amssymb}

\usepackage[german,english]{babel}

\usepackage{authblk}

\newcommand{\be}{\begin{equation}}
\newcommand{\ee}{\end{equation}}

\begin{document}

\title{The Stern-Gerlach Experiment Revisited}

\author[1]{Horst Schmidt-Böcking\thanks{Corresponding author: hsb@atom.uni-frankfurt.de}}
\author[1]{Lothar Schmidt}
\author[2]{Hans Jürgen Lüdde}
\author[1]{Wolfgang Trageser}
\author[3]{Alan Templeton}
\author[4]{Tilman Sauer}

\affil[1]{\small
Institute for Nuclear Physics, Goethe-University Frankfurt, Frankfurt, Germany.} 
\affil[2]{Institute for Theoretical Physics, Goethe-University Frankfurt, Germany.} 
\affil[3]{Oakland, CA, USA.} 
\affil[4]{Institute of Mathematics, Johannes Gutenberg-University Mainz, Mainz, Germany.}  

\date{Version of \today}

\maketitle

\begin{abstract}
The Stern-Gerlach-Experiment (SGE) performed in 1922 is a seminal benchmark experiment of quantum physics providing evidence for several fundamental properties of quantum systems. Based on the knowledge of today we illustrate the different benchmark results of the SGE for the development of modern quantum physics and chemistry. 

The SGE provided the first direct experimental evidence for angular momentum quantization in the quantum world and therefore also for the existence of directional quantization of all angular momenta in the process of measurement. Furthermore, it measured for the first time a ground state property of an atom, it produced for the first time a fully ``spin-polarized'' atomic beam, and it also revealed the electron spin, even though this was not realized at the time. The SGE was the first fully successful molecular beam experiment where the kinematics of particles can be determined with high momentum-resolution by beam measurements in vacuum. This technique provided a kind of new kinematic microscope with which inner atomic or nuclear properties could be investigated. 

Historical facts of the original SGE are described together with early attempts by Einstein, Ehrenfest, Heisenberg, and others to reveal the physical processes creating directional quantization in the SGE. Heisenberg's and Einstein's proposals of an improved multi-stage SGE are presented. The first realization of these proposed experiments by Stern, Phipps, Frisch and Segrè is described. The experimental set-up suggested by Einstein can be considered as an anticipation of a Rabi-apparatus with varying fields. Recent theoretical work by Wennerström and Westlund, by Devereux and others, is mentioned in which the directional quantization process and possible interference effects of the two different spin states are investigated.

In full agreement with the results of the new quantum theory directional quantization appears as a general and universal feature of quantum measurements. One experimental example for such directional quantization in scattering processes is shown. Last not least, the early history of the ``almost'' discovery of the electron spin in the SGE is revisited. 
\end{abstract}

\section{Introduction}

In almost all introductory textbooks on atomic physics the Stern-Gerlach experiment (SGE), performed by Otto Stern (1888--1969) and Walther Gerlach (1889-1979) in Frankfurt in 1922 
\citep{SternO1920Messung,GerlachWEtal1921Nachweis,GerlachWEtal1922Nachweis,GerlachWEtal1924Richtungsquantelung,GerlachW1925Richtungsquantelung}, is presented as a benchmark experiment of quantum science. In most textbooks, the SGE is taken as evidence for proving that Pieter Debye's \citep{DebyeP1916Quantenhypothese} and Arnold Sommerfeld's \citep{SommerfeldA1916Theorie} hypothesis of directional quantization of magnetic and electric momenta of quantum objects in the presence of electric and magnetic fields is a real fact in the quantum world and that magnetic momenta in atoms are quantized. But a more fundamental milestone result, we emphasize, is to be seen in the fact that the SGE provided the first experimental  evidence that, in fact, all angular momenta are quantized in all quantum systems. 

Soon after the advent of the new quantum mechanics, the SGE was recognized as a key experiment to study and understand the problem of measurement in the new theory. As such it was discussed already at the 1927 Solvay conference \cite[esp.~pp.~436, 478]{BacciagaluppiGEtAl2009Crossroads} and by Werner Heisenberg in his paper on the uncertainty relation \citep{HeisenbergW1927Inhalt}. It was then discussed as the paradigmatic example of a theory of the quantum measurement process in David Bohm's textbook on quantum theory \cite[ch.~22]{BohmD1951Quantum}. In recent years, the SGE has been discussed by many authors.

More sophisticated descriptions of the SGE from a physical point of view were presented by \cite{ScullyMOEtal1978Reduction,MackintoshAR1983Experiment,ScullyMOEtal1987Theory,ReinischG1999Experiment}.
Because of its significance as a paradigm for the quantum measurement problem, the SGE has also been discussed both from a historical and a philosophical point of view, see \citep{BernsteinJ2010Experiment}, \citep{SauerT2016Perspectives}. \cite{WeinertF1995Theory} has made the point that the experiment was designed based on a wrong theory, but proved to be the right experiment.

A reconstruction of the historical experiment has been the study of a recent doctoral dissertation \citep{TrageserW2011Effekt}. As a part of this dissertation, the experiment was partially rebuilt, replications of other aspects were reported by \cite{FriedrichBEtal2003Stern,FriedrichBEtal2005Stern}. The historical theoretical context and, in particular, the role of the SGE for the development of the new quantum mechanics was the topic of another recent dissertation \citep{Pie2015Experiment}. It was also discussed from a didactic perspective, e.g.\ by \cite{FrenchAPEtal1978Introduction}, or by \cite{PlattD1992Analysis}.

One aspect, in particular, has received special attention in the literature, the question of identifying mechanisms for the mysterious state reduction of the wave function in the SGE. \cite{DevereuxM2015Reduction} argues that a comparison with double slit experiments of photons or electrons is misleading because in the case of SGE a real energy transfer takes place which destroys any superposition.

The proposal by \cite{FrancaH2009Phenomenon} to explain the alignment of the magnetic moment as a purely classical phenomenon appears not only to be quite ad hoc, it was also criticized for theoretical mistakes by \cite{RibeiroJ2010Was}. More promising seems to be the idea that an early alignment of the magnetic momenta can perhaps be seen as a particular instance of a decoherence process. Thus, 
\cite{GomisP2016Effects} suggest that the alignment is induced by collisions with the remaining gas molecules, an explanation that is challenged, however, by the excellent vacuum conditions of even standard SGEs which produce large mean free paths for the magnetic atoms. 
\cite{WennerstroemHEtal2012experiment,WennerstroemHEtal2013measurements,WennerstroemHEtal2014Interpretation} model the dynamics of the SGE as an interaction between the magnetic moment of the SGE atom with the ensemble of magnetic moments in the SGE magnets. Their model calculations reproduce aspects of the original SGE but have to make non-trivial assumptions that are not fully justified.

Viewing the results of the SGE with the knowledge of today, the SGE provided evidence for the following important milestones in quantum science:   
\begin{enumerate}                                                                                                                                                                                                                                
\item The SGE verified that each silver atom has a magnetic moment of about one Bohr magneton.
\item The SGE presented the first direct experimental evidence that angular momentum is quantized in the quantum world in units proportional to Planck's constant.                       
\item The SGE confirmed  Debye's and Sommerfeld's hypothesis of directional quantization\footnote{Directional quantization (``Richtungsquantisierung'') is a discretization in angle or direction rather than in space. Therefore we will use here the term directional quantization rather than space quantization.}  of magnetic and electric moments of quantum objects in the presence of electric and magnetic fields in the process of measurement.                                                                                                                                                                                                
\item The SGE showed a doublet splitting for silver atoms. As we know today, this splitting is due to the inner magnetic moment of the electron of about one Bohr magneton resulting from the electron spin $= \hbar/2$ with a $g$-factor of about two.                                                                                                                                                           
\item The SGE was the first measurement where a ground-state quantum property of an atom could be determined in a direct way.                                                                                                                                                                                       
\item The SGE produced the first fully spin-polarized atomic beam.                                                                                                    
\item The SGE delivered an atomic beam in a well-aligned state, thus providing the basis for population inversion and therefore proved to be one essential element for the later development of the maser \citep{GordonJEtal1955Maser}.                                                                                        
\item The SGE was the first fully successful molecular beam experiment in high vacuum. Measuring the kinematics of beams and particles with high momentum resolution is tantamount to the invention of a new kind of microscope. Its principle is similar to Aston's mass spectrometer \citep{AstonF1919Spectrograph}. Using an atom or ion beam in a controlled momentum state, the deflection in an outer electric or magnetic field yields information on inner atomic or nuclear properties. 
\item At the time when the SGE was performed, the physics community did not understand why and how the internal magnetic moment (i.e.\ angular momentum) of each atom ``collapses'' in the SG-apparatus into well-defined angular orientations with respect to the direction of the outer magnetic field. This clearly contradicted classical physics where a Larmor precession of the magnetic moments was expected. For most physicists this was a ``miraculous interaction'' between moving atoms and the SG apparatus. With the quantum mechanics of Heisenberg and Schrödinger 
and their new basic equations for describing the dynamical behavior of quantum systems it became clear in a new context that when angular momentum is so small that it becomes comparable to $\hbar$ only well-defined states of directional quantization can exist instead of a continuously varying Larmor precession.
\end{enumerate}

Quantization of angular momentum in units of $\hbar$ is the key element for creating stable dynamical structures in atoms and molecules. The Planck constant $\hbar$ should have the same value in all inertial systems in the universe and it should have no uncertainty as a function of time and location, i.e.\ it should have an ultimate precision. This precision of $\hbar$ is linked to the fact that atoms and molecules cannot emit any extremely soft radiation and should therefore be absolutely stable (Bohr's postulate). Also the dynamics of reactions between atoms, molecules, ions with different projectiles like atoms, molecules, ions, electrons as well as with photons in absorption or emission processes etc.\ is decisively determined by angular momentum conservation and quantization.

To remind the reader of some of the milestones of the emergence of quantum theory,\footnote{For historical accounts of the history of quantum theory, see, e.g.\  \citep{KuhnT1978Theory, MehraJEtAl1982Development1,DarrigolO1992From,KraghH2012Bohr} and further references in these works.} we only mention that quantization of action in quantum systems was discovered by Max Planck (1858--1947). Exploring, theoretically, black body radiation he effectively introduced the novelty that light was quantized (in units now called photons) \citep{PlanckM1899Strahlungsvorgaenge}. A single photon carries the energy $E = h\nu$, where $\nu$ is the frequency of the oscillating photon field and $h$, or $\hbar = h/2\pi$,\footnote{Skipping over historical details that are not in the focus of this paper, we will refer to Planck's constant by its modern symbol $\hbar$ without paying attention to the emergence of this notation and to the fact that, initially, Planck's constant was defined as $h$ rather than $\hbar$.} is a universal constant with the dimension of action or angular momentum. Planck did not yet recognize in 1900 the importance of $\hbar$ for the inner dynamical structure of atoms. Albert Einstein (1879--1955) postulated that photon absorption and emission is also a quantized process where one electron absorbs one photon and vice versa \citep{Einstein1905i}. His simple collision model between photons and electrons delivered thus from the kinematics of the emitted electrons an experimental approach to determine $\hbar$, and explained the physics of the photoelectric effect. In 1911, Otto Sackur (1880--1914) and Hugo Tetrode (1895--1931) discovered, independently from each other, a further important property of $\hbar$ \citep{SackurO1912Anwendung,SackurO1913Bedeutung,TetrodeH1912Konstante1,TetrodeH1912Konstante2}. Defining $\hbar$ as the unit of order in dynamical quantum systems they could calculate the entropy of an ideal gas in an absolute way. In 1913, Niels Bohr (1885--1962) gave $\hbar$ a new dynamical meaning, when he postulated his famous model on the structure of atoms,\footnote{See \citep{BohrN1913Constitution1,BohrN1913Constitution2,BohrN1913Constitution3}, recently reprinted with extensive commentary in  \citep{AaserudFEtal2013Love}.} in which he determined the electronic orbitals in units of the quantized angular momentum $l_z=\hbar m$. Beginning with the Bohr model, $\hbar$ played a crucial role in the forthcoming atomic models. However, the hypothesis of quantization in terms of $\hbar$ was until 1922 only indirectly supported by theory.

The SGE provided a first view into the dynamics of quantum systems and only a few weeks after its results became known \cite{EinsteinAEtAl1922Bemerkungen} were quick to point out that they could not be explained in 1922 by classical understanding of physics. The SGE can be performed for one atom at a time and already in the original SGE the mean free path of the atoms was large enough that silver atoms were effectively travelling without interaction with each other. During the passage of the atom through the SG apparatus one would expect that the dynamical conservation laws of momentum and angular momentum strictly apply yielding well-defined classical trajectories for the atoms. It would only be the orientation of the magnetic moments in well-defined directions when passing through the magnet that contradicts classical physics. The physical mechanism responsible for the alignment of the silver atoms remained and remains a mystery.

Already Stern had pointed out this difficulty in his conceptual paper proposing the SGE:
\begin{quote}
Another difficulty for the quantum conception consists, as has been noted repeatedly, in the fact that we cannot imagine how the atoms of the gas whose angular momenta without magnetic field point in all possible directions, manage to align into the prescribed directions as soon as they are brought into a magnetic field. According to classical theory, one would expect something completely different. Indeed, according to Larmor, the effect of the magnetic field only would be that all atoms begin to rotate uniformly around  the magnetic field as a rotation axis.\footnote{``Eine weitere Schwierigkeit für die Quantenauffassung besteht, wie schon von verschiedenen Seiten bemerkt wurde, darin, daß man sich gar nicht vorstellen kann, wie die Atome des Gases, deren Impulsmomente ohne Magnetfeld alle möglichen Richtungen haben, es fertig bringen, wenn sie in ein Magnetfeld gebracht werden, sich in die vorgeschriebenen Richtungen einzustellen. Nach der klassischen Theorie ist auch etwas ganz anderes zu erwarten. Die Wirkung des Magnetfeldes besteht nach Larmor nur darin, daß alle Atome eine zusätzliche gleichförmige Rotation um die Richtung der magnetischen Feldstärke als Achse ausführen.'' \cite[p.~250]{SternO1921Weg}.}
\end{quote}

After the SGE clearly showed the reality of directional quantization, \cite{EinsteinAEtAl1922Bemerkungen} argued that none of the known physical interactions could account for the alignment of the atoms. The problem has remained a puzzle to many physicists ever since. 
Thus, Julian Schwinger (1918--1994) said in his lecture book on quantum mechanics:
\begin{quote}
It is as though the atoms emerging from the oven have already sensed the direction of the field of the magnet and have lined up accordingly. Of course, if you believe that, there's nothing I can do for you. No, we must accept this outcome as an irreducible fact of life and learn to live with it! 
\cite[p.~30]{SchwingerJ2001Mechanics}
\end{quote}
Richard Feynman (1918--1988) wrote in his lectures: 
\begin{quote}
That a beam of atoms whose spins would apparently be randomly oriented gets split up into separate beams is most miraculous. How does the magnetic moment know that it is only allowed to take on certain components in the direction of the magnetic field? Well that was really the beginning of the discovery of the quantization of angular momentum, and instead of trying to give you a theoretical explanation, we will just say that you are stuck with the result when the experiment was done.  
\cite[Vol.II, 35-2]{FeynmanR1963Lectures}
\end{quote}
According to Stern, Einstein, Ehrenfest, and later Schwinger, Feynman and others, the unsolved puzzle was: Why and where in the apparatus could the passing atoms interact with the SG-apparatus in a way that their magnetic moment point into certain directions with respect to the direction of the outer magnetic field $B$?  

We know today that the directional quantization of angular momentum and magnetic moments in magnetic fields as observed in the Zeeman Effect \citep{ZeemanP1896Invloed,ZeemanP1897Influence} as well as in the Stern-Gerlach experiment are closely related processes. Thus it is astonishing why scientists in 1922 accepted the Zeeman Effect as a new signature of quantum physics and vice versa found the SGE observations partly ``miraculous''. Pieter Zeeman's (1865--1943) apparatus and the SG apparatus share some common features in creating directional quantization. 

But they are also clearly different in the detection approach. In both experiments, the atomic magnetic momenta are directionally quantized with respect to the magnetic field $B$. In accordance with the Maxwell velocity distribution, the atoms in Zeeman's experiment move randomly in all directions with respect to the $B$ field, colliding frequently with the other gas atoms. Zeeman detected only photons emitted from excited states. From the photon energies $\Delta E \propto B\cdot \mu\cos\alpha$ the angles $\alpha$ of orientation of the magnetic moments $\mu$ in the $B$ field (i.e.\ directional quantization) were determined. But the atoms themselves were not detected. 

In Stern's apparatus, the atoms were evaporated in an oven and then injected into the vacuum. By setting narrow slits a beam of atoms with a well-defined momentum $p_x$ in $x$-direction was produced and injected into the magnetic field $B$. Because of high vacuum conditions the beam atoms did not undergo any further collision with the rest gas molecules in the vacuum. Passing the $B$ field in the entrance region the magnetic moment of the silver atoms became directionally quantized like the atoms in the Zeeman Effect. Because of their momentum $p_x$ and the high vacuum in the SGE the atoms pass through the entrance region (duration of a few micro seconds) and enter the inhomogeneous $B$ field region inside the magnet. 

In Stern's apparatus the atoms were prepared into a fully controlled dynamical momentum state. By measuring the dynamical properties (momentum), perfect control of directional quantization along the atom's trajectory inside the magnetic field region was obtained. After entering the inhomogeneous $B$ field region between the poles of the magnet each atom was accelerated by the magnetic force $\partial B/\partial z\cdot \mu$ in $z$-direction due to its magnetic moment $\mu$. At the point of exiting the magnetic field region each atom has a well-defined transverse momentum $\Delta p_z\propto (\partial B/\partial z) \cdot \mu \cdot t_F$ ($t_F$ is the transit time of the atom inside the magnetic field). Measuring this transverse momentum, which was found to have two discrete values, allowed Stern and Gerlach to determine the value of the quantized magnetic moment. It is to be noticed that the SGE really provides a momentum measurement and not a measurement of position.  Each atom appears to follow a perfectly steady, classical trajectory in the SG device starting from the oven until after leaving the magnetic field. Since the de Broglie wave length $\lambda$ is $< 0.02$~\AA, diffraction at the slits is completely negligible. 

\begin{figure}
\begin{center}
\includegraphics[scale=0.5]{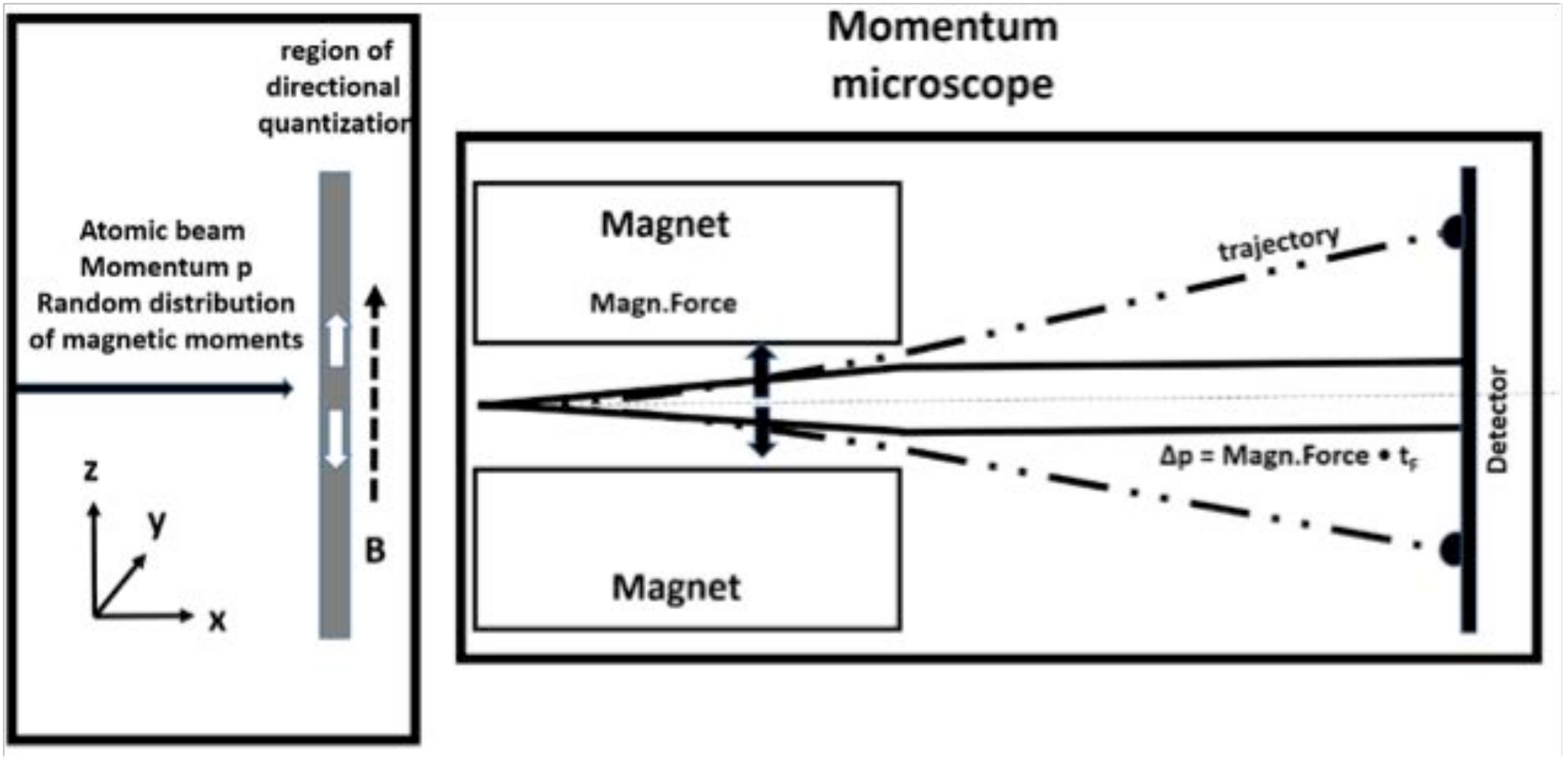}
\caption{Conceptual scheme of Stern's momentum microscope to reveal directional quantization. The left box shows the very short region, where directional quantization is achieved in the $B$-field, and the right box shows the momentum microscope design where by an inhomogeneous magnetic force $\partial B/\partial z$ the different orientations of the magnetic moments $\mu_L$ of the moving beam are dynamically separated (dashed-dotted lines represent the trajectories; the solid line represents the transverse momentum $\Delta p$).}
\label{fig:SGEscheme}
\end{center}
\end{figure}

In Figure \ref{fig:SGEscheme}, the scheme of Stern's momentum microscope as realized in the SGE is illustrated. Conceptually, it consists of two parts: on the left hand side it shows a region, where directional quantization is achieved (non-classical interaction) in the $B$-field (in all probability a very small region at the entrance of the magnet), and on the right hand side it shows the momentum microscope design where the different orientations of the magnetic moments $\mu_L$ of the moving beam are dynamically separated by an inhomogeneous magnetic force $\partial B/\partial z$. The microscope part on the right is a purely classical apparatus which transfers the different magnetic momenta into different transverse momentum states $\Delta p_z$. It works in a quite similar way as a mass spectrometer, where different charge states or masses are deflected in an electric or magneto-static field into different angles (transverse momenta). 

In 1919, Otto Stern had established the foundations for this microscope in Frankfurt with the development of the molecular beam method (MBM) \citep{SternO1920Messung}. This happened at about the same time when Arthur Jeffrey Dempster (1886--1950) and Francis William Aston (1877--1945) developed their ion beam mass spectrographs \citep{DempsterAJ1918Method,AstonF1919Spectrograph}. Stern was the first to prepare beams of single isolated atoms in a vacuum with controlled velocity and direction. Thus he was able to measure transverse-momentum transfers with excellent resolution when the atoms were deflected by external forces like in the SGE, with a momentum resolution of ca. 0.15~a.u.\ (``atomic units'' in which we set $e=m_e=\hbar=1/(4\pi\epsilon)=1$. An electron of $13.6$~eV kinetic energy has a momentum of 1 a.u.). 
 
The trajectories of the atoms can be calculated using equations of motion from classical physics, but the rotation of the magnetic moments into well-defined orientations remain a puzzle. A few weeks after the SGE had successfully been performed in 1922, Albert Einstein and Paul Ehrenfest considered all possible interactions of momentum exchange by any kind of classical forces and of radiation exchange between atom and apparatus to explain this rotation \citep{EinsteinAEtAl1922Bemerkungen,UnnaIEtal2013Einstein}. They came to the conclusion that radiation exchange would take more than $100$ years to turn the angle. The ``puzzle'' of directional quantization, to explain how atom and SG apparatus ``interact which each other'', could not be solved by Einstein and Ehrenfest in 1922. 
 
In Einstein's and Ehrenfest's analysis, the only force active in the SGE was the magnetic force. According to classical physics it should induce a Larmor precession of the magnetic moment around the $B$-field vector. This Larmor precession would induce additional Larmor radiation but quantitatively Einstein and Ehrenfest found that this process would take place on time scales many orders of magnitude larger than that set by the time of flight through the magnetic field region. Therefore it should not change measurably the angle relative to the $B$ field. In classical physics, where the angular momentum vectors are huge compared to $\hbar$ one assumes that Larmor precession is a process continuous in angle with respect to the outer magnetic field vector. But also Larmor precession is quantized in units of $\hbar$. Only when the total angular momentum becomes rather small and approaches $\hbar$, i.e.\ when experiments on single atoms or ions are performed this quantization of the Larmor precession becomes visible and its components on the outer field direction must be multiples of $\hbar$, too. Only in recent years, \cite{HermansphanNEtal2000Observation} have shown that the Larmor precession of ions moving in a trap (a continuous Stern-Gerlach-like apparatus) is indeed quantized.
 
We know today that directional quantization of angular momenta and thus of magnetic moments is always present and visible in the structure and dynamics (and reactions too) of atomic and molecular systems. In the new quantum theory of matrix- and wave mechanics, angular momentum appeared quantized in length and direction in clear contradiction with classical physics.
 
One example of a recent SGE-like measurement where directional quantization of angular momenta is observable will be presented in section 5: fully-differential data on the single electron emission in slow He$^{2+}$-on-He scattering experiments (quasi an SGE of electrons in the electric field of the $\alpha$-$\alpha$ nucleus quasi molecule) \citep{SchmidtLEtal2014Vortices}. As shown below, the momentum distribution of the emitted electrons and deflected atoms is always completely determined by angular momentum exchange and its directional quantization. In any reaction or transition in quantum systems angular momentum is always exchanged in quantized values determined by $\hbar$. Thus all quantum dynamics is ``discretized'' because of the finite value of $\hbar$.   
 
For Heisenberg and Einstein the SGE was a seminal experiment with benchmark results. Thus it is not surprising that soon after the SGE was performed, both proposed nearly identical suggestions of an improved multi-stage SGE to explore more secrets of the directional quantization process of the magnetic moments. These proposed experiments of an improved multi-stage SGE and the realization of these proposals by Thomas Erwin Phipps (1896--1990) and Otto Stern, and by Robert Otto Frisch (1904--1979) and Emilio Segrè (1905--1989)  are revisited \citep{PhippsTEtal1932Einstellung,FrischOEtal1933Einstellung}. Heisenberg's and Einstein's proposed experimental set-up can be considered a pre-Rabi apparatus with varying fields. 

\cite{WennerstroemHEtal2012experiment,WennerstroemHEtal2013measurements,WennerstroemHEtal2014Interpretation} and Michael \cite{DevereuxM2015Reduction} have explored the passage of single atoms through an SG apparatus. In both theoretical approaches the process of directional quantization is investigated taking into account the dynamical coupling of a single atom (momentum and angular momentum conservation) with the SG apparatus. \cite{WennerstroemHEtal2012experiment,WennerstroemHEtal2013measurements,WennerstroemHEtal2014Interpretation} consider a stochastic coupling of the atomic magnetic moment to the spins of the atoms in the magnet.  \cite{DevereuxM2015Reduction} investigates the passage of single atoms through the magnet and shows that the two separated spin states of the silver beam cannot interfere since their trajectories are experimentally distinguishable.
 
Last not least in chapter 6 the SGE will be revisited with respect to the history of electron spin discovery. 
 
\section{Remarks on the historical SGE} 
 
Roughly, the collaboration between Stern and Gerlach can be characterized by saying that Otto Stern delivered the concept and design of the SGE and Walther Gerlach made it work.%
\footnote{For a detailed historical account of the SGE, see \citep{TrageserW2011Effekt}.}
\begin{figure}
\begin{center}
\includegraphics[scale=0.5]{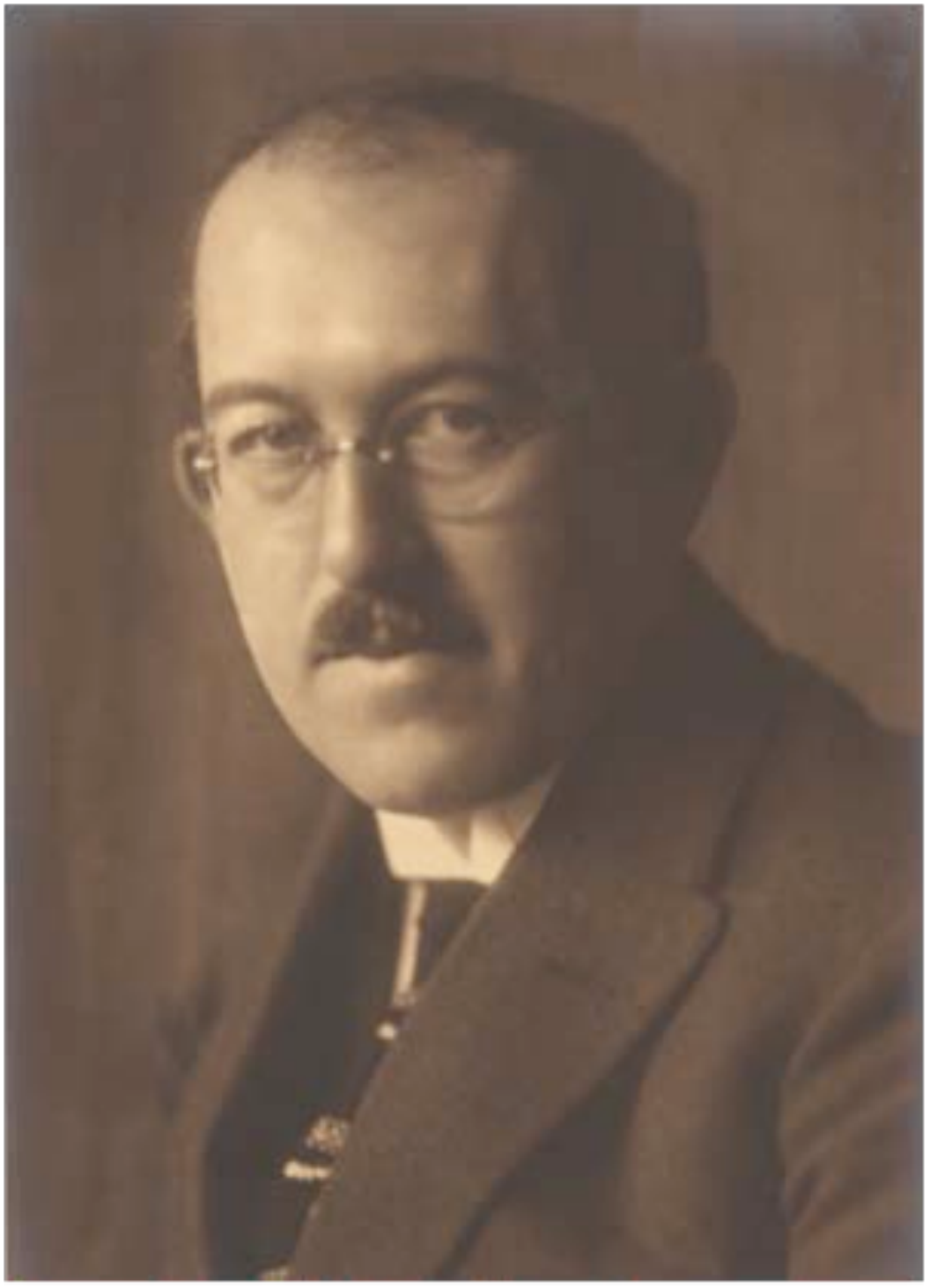}
\includegraphics[scale=0.505]{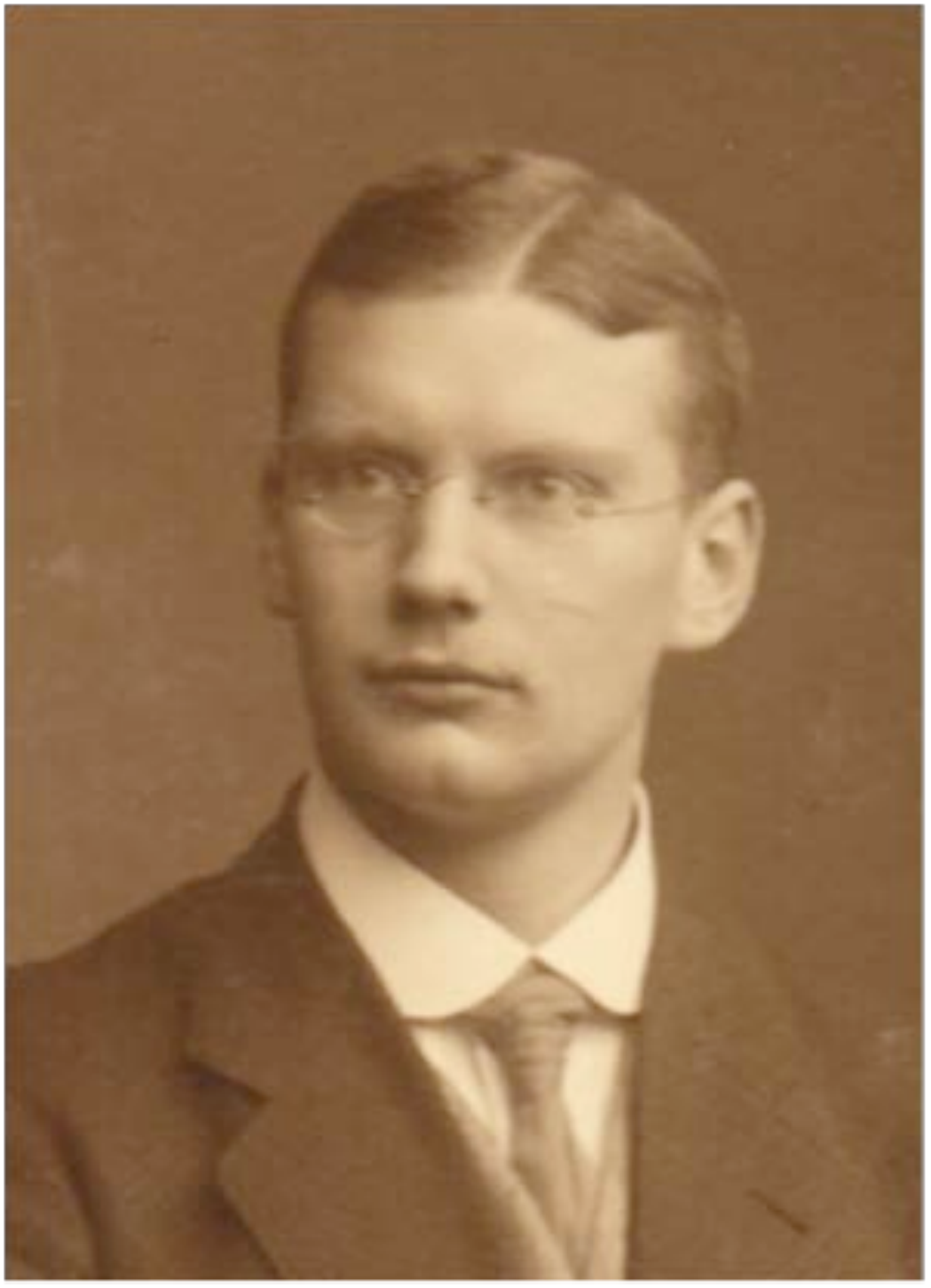}
\caption{Otto Stern 1920 and Walther Gerlach 1911 (Picture: US-OSF; Gy-UFA, donated by Werner Kittel).}
\label{fig:SternGerlach}
\end{center}
\end{figure}
After Stern had finished his doctoral dissertation on a topic in Physical Chemistry at the university of Breslau (present-day Wroc\l{}aw) in April 1912, he became Albert Einstein's assistant at the German Charles University in Prague.\footnote{For Otto Stern's biography, see  \citep{SchmidtBoeckingHEtal2011Stern,ToenniesJEtal2011Stern}, see also Sz-ETHA, Otto Stern tape-recording, Folder ``ST-Misc'', 1961 at E.T.H.\ Z\"urich by Res Jost; US-OSF; US-NBLA, Interview with Dr.~Otto Stern, by Thomas S.~Kuhn at Stern's Berkeley home, May 29 \& 30, 1962. Most of Otto Stern's papers are located at US-BK, BANC MSS 85/96 c.
See also \citep{FriedrichBEtal1998Star}.
\label{note:stern_bio}} In Oktober 1912, he followed Einstein and moved to Zurich to work at the ETH. In 1913, when Einstein accepted an offer to become a member of the Prussian Academy as well as director of a ``Kaiser Wilhelm Institut für Physik'' in Berlin, Stern accepted Max von Laue's (1879--1960) offer to come to Frankfurt as a Lecturer (``Privatdozent'') in theoretical physics at its newly (1914) founded “Royal University” (``Königliche Universität Frankfurt''). Although Otto Stern had been a ``Privatdozent'' in Frankfurt since November 1914, he volunteered for the army when the war started and served as a weather observer in Lomsha, East Poland. At the end of the Great War, after a research stay of several months in Walter Nernst's institute in Berlin, Stern returned to the University of Frankfurt in February 1919. 

When Otto Stern heard for the first time about ``Richtungsquantelung'' in a seminar in 1919, he was convinced that the idea was nonsense.\footnote{As expressed in his Zurich interview, see the references in the previous footnote.} For him it was clear that the idea implied that light passing through gaseous matter in a magnetic field should exhibit diffraction and that somebody surely would have observed the effect if it were real. However, it was characteristic of Stern's attitude that he immediately recognized that he could put this quite unbelievable, but important hypothesis to experimental test with his newly developed molecular beam method.  

Together with Walther Gerlach, Stern began work on the SGE already in 1920 based on his molecular beam method \citep{SternO1920Messung,SternO1920Messung2,SternO1920Nachtrag}. In 1921, Stern published his idea of the SGE in a single-authored paper in \emph{Zeitschrift für Physik} \citep{SternO1921Weg}. It is remarkable that Stern was trained as a theoretical physicist and a chemist. He became Professor (on a non-tenured position) for theoretical physics at the University Frankfurt in 1919 when Max Born (1882--1970) was director of the institute. 

Walther Gerlach obtained his PhD in experimental physics at the institute of Friedrich Paschen in Tübingen in 1912. He also habilitated there in 1916.\footnote{For Walther Gerlach's biography, see the interview with Dr.~Walther Gerlach by Thomas S.\ Kuhn at Gerlach's home, Berlin, West Germany, February 18, 1963 US-NBLA; and GyDM, Nachlass Walther Gerlach. See also \citep{HuberJG2014Gerlach,FuesslW1998Nachlass,HeinrichREtal1989Gerlach, NidaRuemelinM1982Bibliographie}.} After short stays at the University of Göttingen as a ``Privatdozent'' and in industry he accepted the position of an assistant in the institute of experimental physics (directed by Richard Wachsmuth (1868--1914)) at the University of Frankfurt in 1920. In 1919, shortly after the Great War, research activities were greatly hampered by missing resources \citep{FrickeHNoDateJahre}. 

Luckily, the institute of theoretical physics in Frankfurt owned a mechanical workshop, and in it a  capable young mechanic, the ``Mechanikermeister'' Adolf Schmidt (1893--1971) put his talents to good use. Stern and Born began to use the resources of the mechanical workshop to perform experiments. Stern, in particular, used the workshop exceptionally well. Within a few months in 1919 he laid the foundations of the molecular beam method starting with a direct measurement of the Maxwell velocity distribution of evaporated Silver atoms at a given temperature T \citep{SternO1920Messung,SternO1920Messung2,SternO1920Nachtrag}. Stern designed experiments in his ingenious way, Adolf Schmidt manufactured the parts of the experiments in a nearly perfect way, and Walther Gerlach was able to mount them all together and made the apparatus really work. Born helped Stern and Gerlach to raise money (Born gave public lectures on Einsteins ``Theory of Relativity'' (published as \citep{BornM1920Relativitaetstheorie}) and contacted Henry Goldman in New York, the former CEO of Goldman \& Sachs; Gerlach obtained money from Einstein as director of the Berlin Kaiser Wilhelm institute;\footnote{Walther Gerlach wrote on July 27$^{\rm th}$, 1921, a letter to Einstein as director of the ``Kaiser Wilhelm Institut für Physik'' in Berlin asking for support (request 8000RM) \cite[p.~783]{CPAE13}. This request was approved on December 2$^{\rm nd}$, 1921, 
and they received 10000~RM \cite[pp.~476, 477, 479, 482]{CPAE12}.} they were also getting material support like magnets, liquid air for cooling etc.\ from local companies in Frankfurt \citep{SchmidtBoeckingHEtal2011Stern,ToenniesJEtal2011Stern}). Considering the circumstances, it was a remarkable achievement that such a difficult experiment could be performed immediately after the war and at a time of beginning monetary ``inflation''. 
 
\begin{figure}
\begin{center}
\includegraphics[scale=0.5]{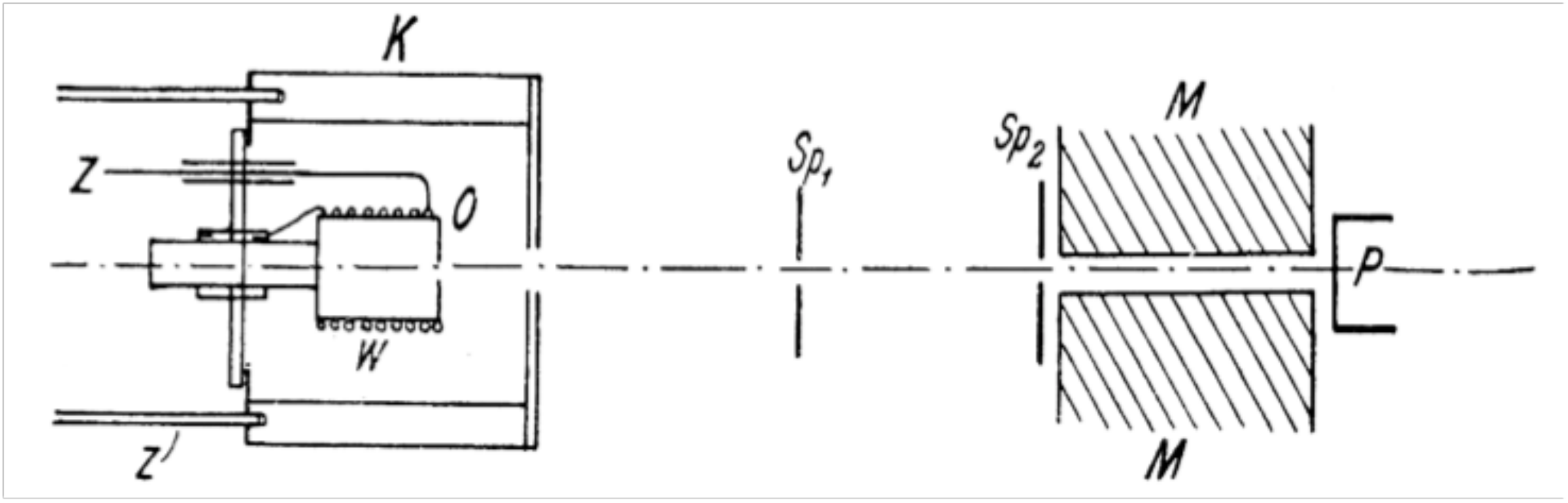}
\caption{Scheme of the original SGE \cite[p.~677]{GerlachWEtal1924Richtungsquantelung}. The silver is heated in the furnace which is electrically heated by a wire W with electric leads Z. The oven is cooled by a cooler K, Sp1 and Sp2 are collimators, M is the magnet and P a cooled detection plate (Picture from Stern's private slide collection, Gy-UBFAZ).}
\label{fig:SGEorigscheme}
\end{center}
\end{figure}

The successful performance of the SG experiment was put at risk 1921 when Born received an offer for a professorship in Göttingen which could have meant that all experiments in Born's institute might have come to an end in Frankfurt.\footnote{For Born's call to G\"ottingen and the subsequent negotiations, see the documents in Gy-UFA, PA Max Born, see also \cite[esp.~Docs.~75, 95]{CPAE10} and \citep{DahmsHJ2002Politics}.}
In his negotations, Born expressed his willingness to remain in Frankfurt. One of his five conditions for staying was that Stern was offered a permanent professorship in Frankfurt. But this requirement was denied by the University of Frankfurt, and Born accepted the offer from Göttingen. Therefore, the failure to make any efforts to keep Stern in Frankfurt was also one of the reasons that Born left the university. Stern, too, left Frankfurt in October 1921, accepting a professorship for theoretical physics at the University of Rostock. Gerlach continued to work on the SGE, in close scientific exchange with Stern. 

The SG apparatus finally got to a stage when it was producing useful results in fall 1921 just at the time when Stern was leaving Frankfurt. On November 4$^{\rm th}$, 1921, Gerlach saw the first broadening of the silver beam spot (black silver sulfide) on the detector plate, when the magnetic field was on \citep{GerlachWEtal1921Nachweis,GerlachW1969Stern,GerlachW1969Entdeckung}. Since the resolution then was still very low the spot structure allowed only a rough estimate of the magnitude of the silver atom's magnetic moment. But it already showed that the magnetic moment was roughly of the expected size of one Bohr magneton. In this first experiment, the resolution did not yet allow Gerlach to see a splitting of the atomic beam \citep{GerlachWEtal1921Nachweis}.    

\begin{figure}
\begin{center}
\includegraphics[scale=0.45]{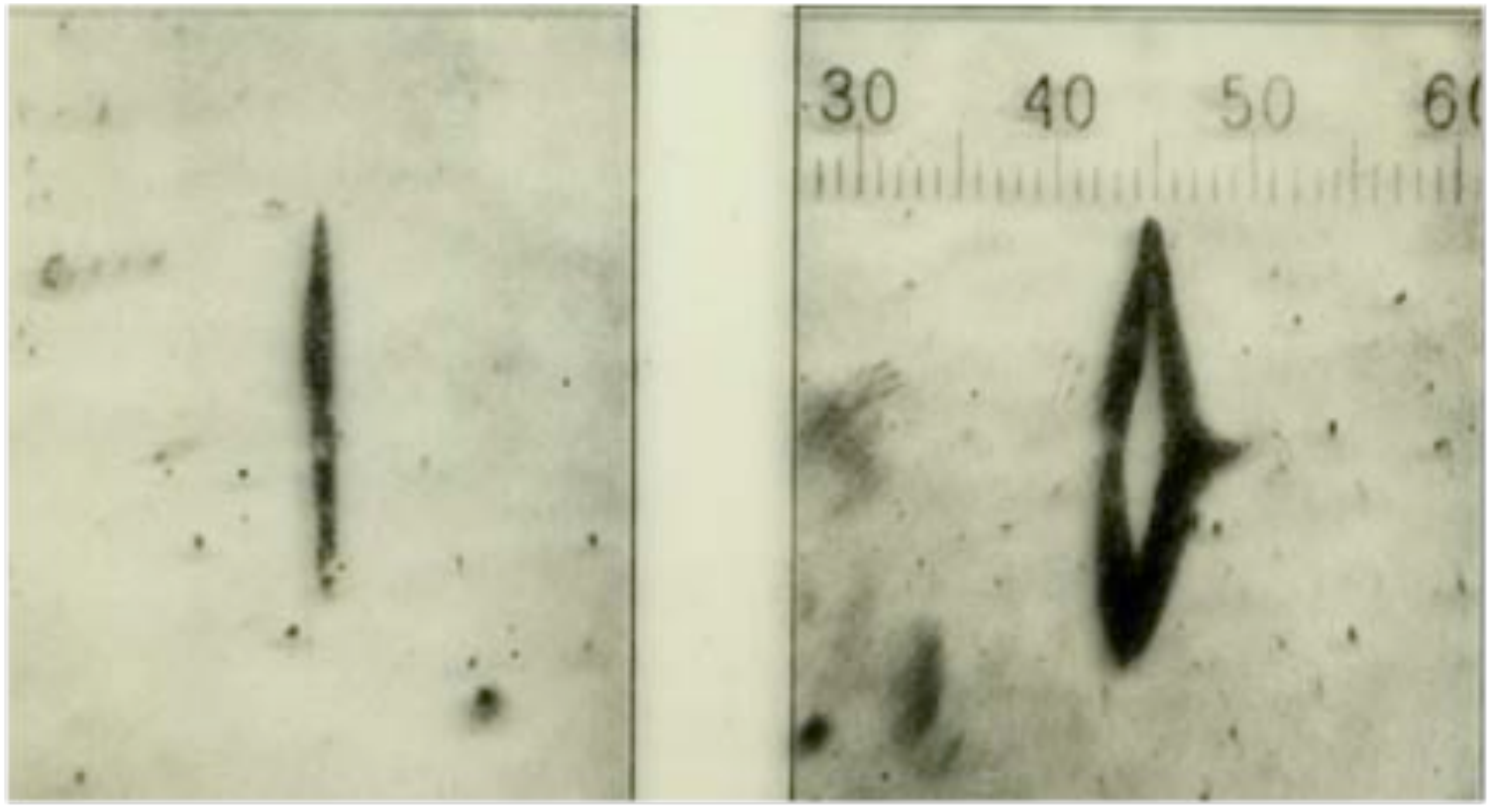}
\includegraphics[scale=0.45]{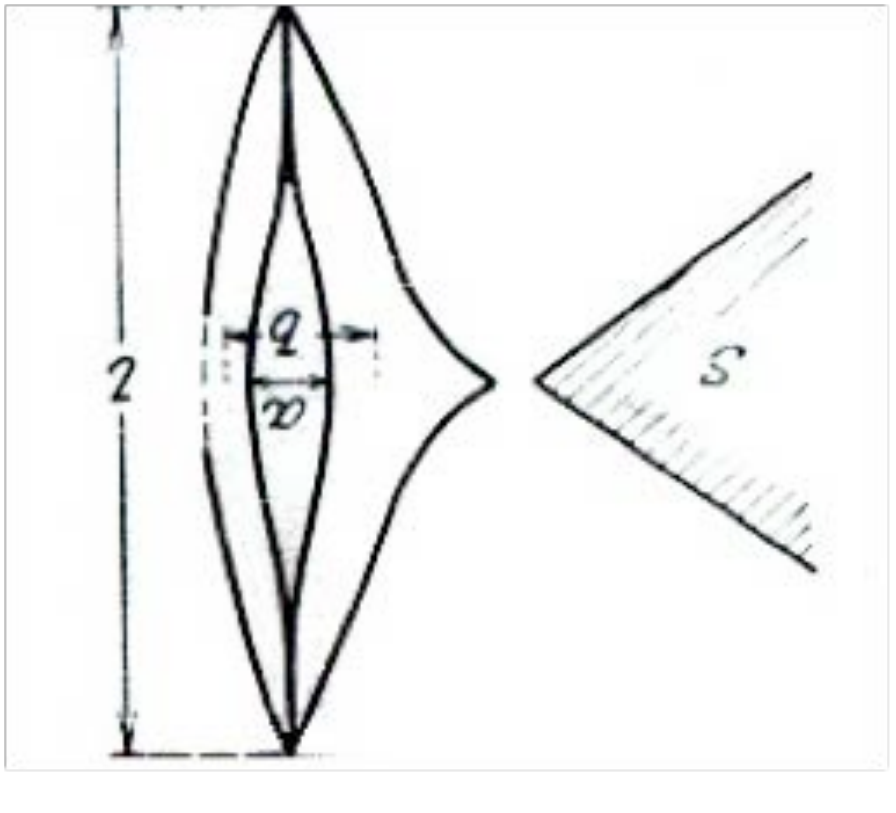}
\caption{Observed pattern on the detector plate: left without magnetic field, middle with magnetic field and right beam spot geometry near the edge of the magnet. Since the magnetic field strength is fast decreasing with distance from the edge of the magnet (perpendicular to the direction of the $B$-Field) the beam components merge. (Images from Stern's private slide collection, Gy-UBFAZ, see also \cite[pp.~350, 351]{GerlachWEtal1922Nachweis})}
\label{fig:SGEobservedpattern}
\end{center}
\end{figure}

One of the crucial parameters of the first SGE design was the collimation of the silver beam \citep{GerlachWEtal1922Nachweis}. The silver beam had to pass first through a tiny oven aperture ($1$~mm diameter), then in $2.5$~cm distance through a nearly circular aperture (area of $3\cdot10^{-3}$~mm$^2$) and then just before the entrance into the magnetic field (at a distance of $3.5$~cm from the second aperture) it passed through a rectangular aperture of $0.8$mm length and $0.3$--$0.4$~mm width. It was nearly impossible to get a controlled beam with enough intensity passing through all three holes. In the first days of February 1922, Stern and Gerlach met at a small conference in Göttingen \citep{FriedrichBEtal2003Stern} and decided to exchange one circular aperture by a rectangular slit. This modification proved immediately to be the right thing to do. A few days later, in the night from February $7^{\rm th}$ to $8^{\rm th}$, 1922, the SGE was successful \citep{GerlachWEtal1922Nachweis}.\footnote{See also Gerlach's typescript ``Die entscheidenden Stufen f\"ur den Nachweis der Richtungsquantelung,'' dated 22 February, 1963, 3pp., Gy-DM, NL080, Nachlieferung.}  See Figure \ref{fig:SGEorigscheme} for a sketch of the eventually successful SGE setup. Being a night worker, only Gerlach had supervised the experiment this night. A PhD student Wilhelm Schütz joined him in the morning of February $8^{\rm th}$ \citep{SchuetzW1969Erinnerungen}. Stern was already back in Rostock. Because of the small size of the SG apparatus (about $3$~cm distance between collimators) the observed doublet structure was barely separable (about $0.1$~mm) but a microscopic photograph of the detector plate clearly showed a distinct separation of two beams, see Figure \ref{fig:SGEobservedpattern}.

The result was in clear contradiction to the outcome predicted according to the classical theory. Classically, only a broadening of the observed spot should be expected. According to quantum theory, the beam should split up but there was some disagreement whether it should split up into two or three beams. Sommerfeld's quantization hypothesis required that the angular momentum vector of the outer electron orbit should be aligned parallel, orthogonal, or anti-parallel to the direction of the magnetic field. If all three directions were possible in nature, one would see a \emph{triplet} splitting with an undeflected central beam. Bohr, on the other hand, had postulated that a quantization with the angular momentum orthogonal to the field was dynamically impossible. He therefore predicted a \emph{doublet} splitting. Arnold Sommerfeld's position at the time is unclear but he probably \citep{SommerfeldA1920Zahlenmysterium,SommerfeldA1920Gesetze}, \cite[p.~541]{SommerfeldA1921Atombau2} had predicted a triplet splitting assuming a magnetic moment of one magneton and a splitting into one component up, one down and one perpendicular like in the normal Zeeman effect (triplett structure). 

Since a clear doublet splitting was observed (Figure~\ref{fig:SGEobservedpattern}), it was believed that Niels Bohr was right and nobody debated his explanation. Gerlach immediately sent him a postcard with a photo of the result on February $8^{\rm th}$ stating stating that Bohr's theory had been confirmed \citep{GerlachW1969Stern,GerlachW1969Entdeckung,FriedrichBEtal2003Stern}. Although in a recent doctoral thesis, it is argued that the immediate reception of the SGE did not play a major role for the development of quantum theory \citep{Pie2015Experiment}, there is some evidence that the SGE results did have some impact on the physics community.\footnote{See the references in footnote \ref{note:stern_bio} and \citep{GerlachW1969Stern}.} 

In the fourth edition of \emph{Atombau und Spektrallinien}, Arnold Sommerfeld acknowledged:
\begin{quote}
By their bold experimental design, Stern and Gerlach not only demonstrated ad oculos the spatial quantization of atoms in a magnetic field but they also proved the atomistic nature of the magnetic moment, its quantum theoretical origin, and its relation to the atomistic structure of electricity.\footnote{``Durch ihre kühne Versuchsanordnung haben also Stern und Gerlach nicht nur die räumliche Quantelung der Atome im Magnetfelde ad oculos demonstriert, sondern sie haben auch die atomistische Natur des magnetischen Momentes, seinen quantentheoretischen Ursprung und seinen Zusammenhang mit der atomistischen Struktur der Elektrizität bewiesen.''
\cite[p.~149]{SommerfeldA1924Atombau4}.}
\end{quote}
Albert Einstein wrote in May 1922:
\begin{quote}
The experiment by Stern and Gerlach is the most interesting at the present time, though. The atoms' orientation without collisions cannot be explained by radiation (according to our current methods of considering the problem). By rights, an orientation ought to persist longer than 100 years.\footnote{``Das Interessanteste aber ist gegenwärtig das Experiment von Stern und Gerlach. Die Einstellung  der Atome ohne Zusammenstösse ist nach (den jetzigen Überlegungs Methoden) durch Strahlung nicht zu verstehen. [-] eine Einstellung sollte von Rechts-Wegen mehr als 100 Jahre dauern.''  Einstein to Max Born, on or after 14 May 1922 \cite[Doc.~190]{CPAE13}.}
\end{quote}
And Wolfgang Pauli (1900--1958) wrote in a postcard to Gerlach: 
\begin{quote}
Now, hopefully also the unbelieving Stern will be convinced of directional quantization.\footnote{``Jetzt wird hoffentlich auch der ungläubige Stern von der Richtungsquantelung überzeugt sein.“ 
A.~Sommerfeld to W.~Gerlach, 17~February 1922, \cite[p.~55]{PauliW1979Briefwechsel}.}
\end{quote}
Even long after the success of the SGE, Stern had always problems to accept the physics hidden in the SGE results. In an interview with Jost Fierz conducted in Zurich in 1961, Stern still expressed doubts about the interpretation of the SGE results. He said: 
\begin{quote}
But with the outcome of the experiment, I really did not understand anything... It was absolutely unintelligible. But this is very clear since you need not only the new quantum theory but also the magnetic electron. These two things that weren't there at the time. I was totally confused and did not know at all what to make of it. I still have objections against the beauty of quantum mechanics. But it is a correct theory.\footnote{``Aber wie nun das Experiment ausfiel, da hab ich erst recht nichts verstanden. .... Das war absolut nicht zu verstehen. Das ist auch ganz klar, dazu braucht man nicht nur die neue Quantentheorie, sondern gleichzeitig auch das magnetische Elektron. Diese zwei Sachen, die damals noch nicht da waren. Ich war völlig verwirrt und wusste gar nicht, was man damit anfangen sollte. Ich habe jetzt noch Einwände gegen die Schönheit der Quantenmechanik. Sie ist aber richtig.'' Interview with Stern, see note \ref{note:stern_bio}.}
\end{quote}

After their successful performance of the SGE, Stern and Gerlach received the highest international reputation in physics. According to the official listing in the Nobel archives  (Sw-RSAS), Stern and Gerlach were nominated as a duo, beginning in 1925, 31 times for the Nobel Prize in physics. The last nomination for both physicists together came from Manne Siegbahn in 1944 who was at that time the chairman of the Nobel committee for physics.  Stern alone received an additional 51 nominations and was nominated a total of 82 times.\footnote{The official number of nomination in the Nobel archives (The Nobel Population 1901-1950, A census 2002 The Royal Swedish Academy, Produced by Universal Academy Press, Inc.) 
has 81 nominations for Otto Stern. In the official list, Gregor Wentzel is  listed only four times as nominator for Stern (for the years 1938, 1940, 1941, and 1944). His letter of nomination for Otto Stern of 5 January 1943 was not taken into account, although the letter is extant in the Novel archives. Also, Viktor Hess claims in his letter to Otto Stern (Stern papers in the Bancroft Archives microfilm Nr. 69~0133) that he nominated Stern in the years 1937 and 1938 for the Nobel prize in physics. According to the curator of the Nobel archives, Karl Grandin, this claim is false.}

The first nomination for Stern and Gerlach came from Einstein in 1923 for the year 1924 \cite[Doc.~132]{CPAE14}. This nomination is not listed in the official listing since Einstein nominated in this letter several candidates and ranked Franck and Hertz on first place. Reading the nomination letters from most of the nominators (e.g. James Franck, Max Born, Max von Laue, Max Planck, Albert Einstein, Niels Bohr, Hans Bethe, Oskar Klein, Werner Heisenberg, Eugen Wigner, Carl David Anderson, Wolfgang Pauli etc.) it is quite clear that Stern was considered the ``spiritus rector'' (guiding spirit) behind the SGE. No one, in fact, challenged Einstein's and Ehrenfest's early referring to the experiment as the experiment by Stern and Gerlach. Although all their joint publications on the SGE were authored alphabetically \citep{GerlachWEtal1921Nachweis,GerlachWEtal1922Moment,GerlachWEtal1922Nachweis,GerlachWEtal1924Richtungsquantelung}, they were predated by Stern's single-authored proposal of the method \citep{SternO1921Weg}.

\begin{figure}
\begin{center}
\includegraphics[scale=0.5]{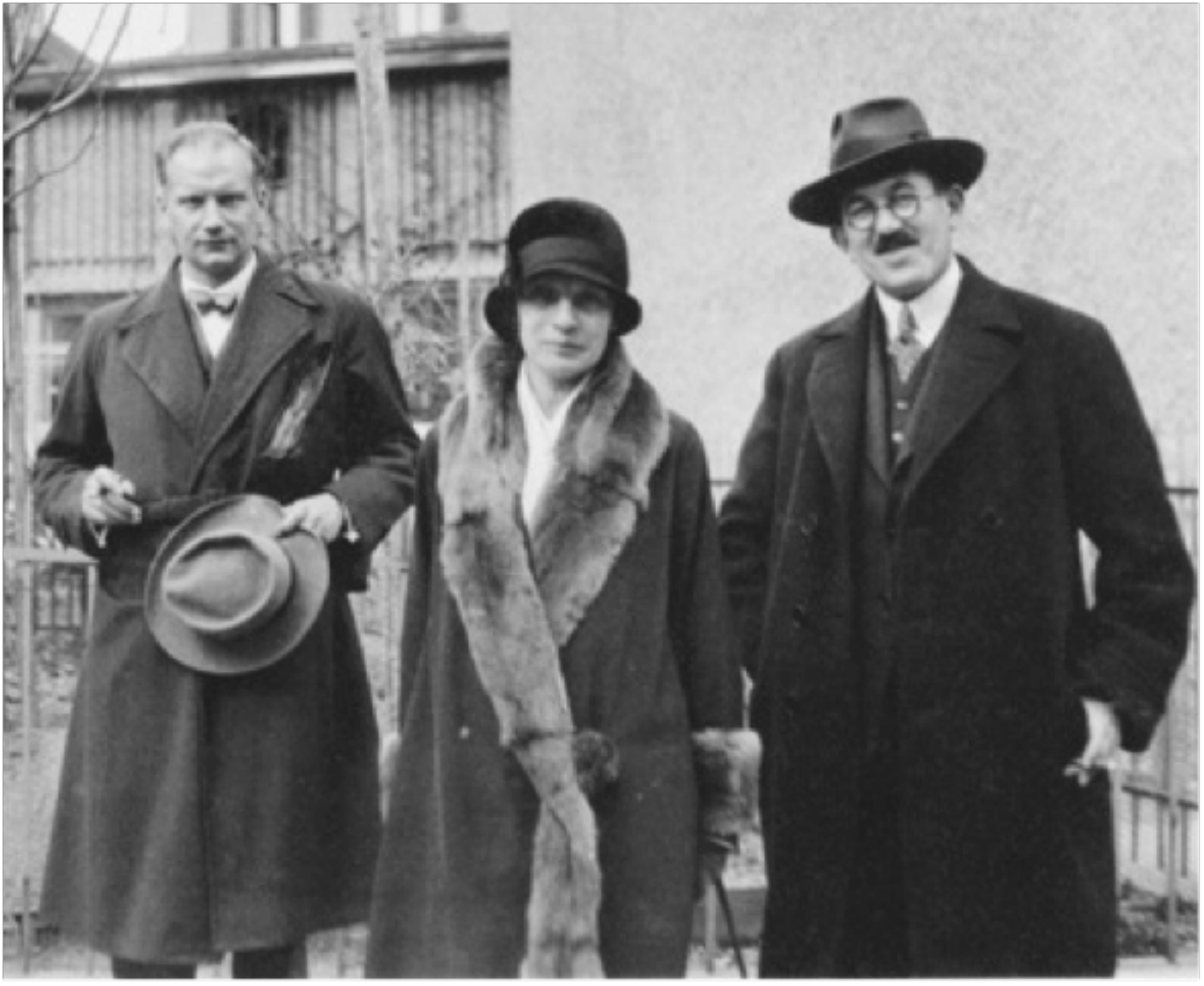}
\caption{About 1927 in Zürich. From right: Otto Stern, Lise Meitner, and Walther Gerlach (Picture collection of Ruth Speiser-Bär (``Ellen Weyl-Bär, Privatbesitz'') and Bruno Lüthi,  private communication.)}
\label{fig:SternMeitnerGerlach}
\end{center}
\end{figure}
\begin{figure}
\begin{center}
\includegraphics[scale=0.45]{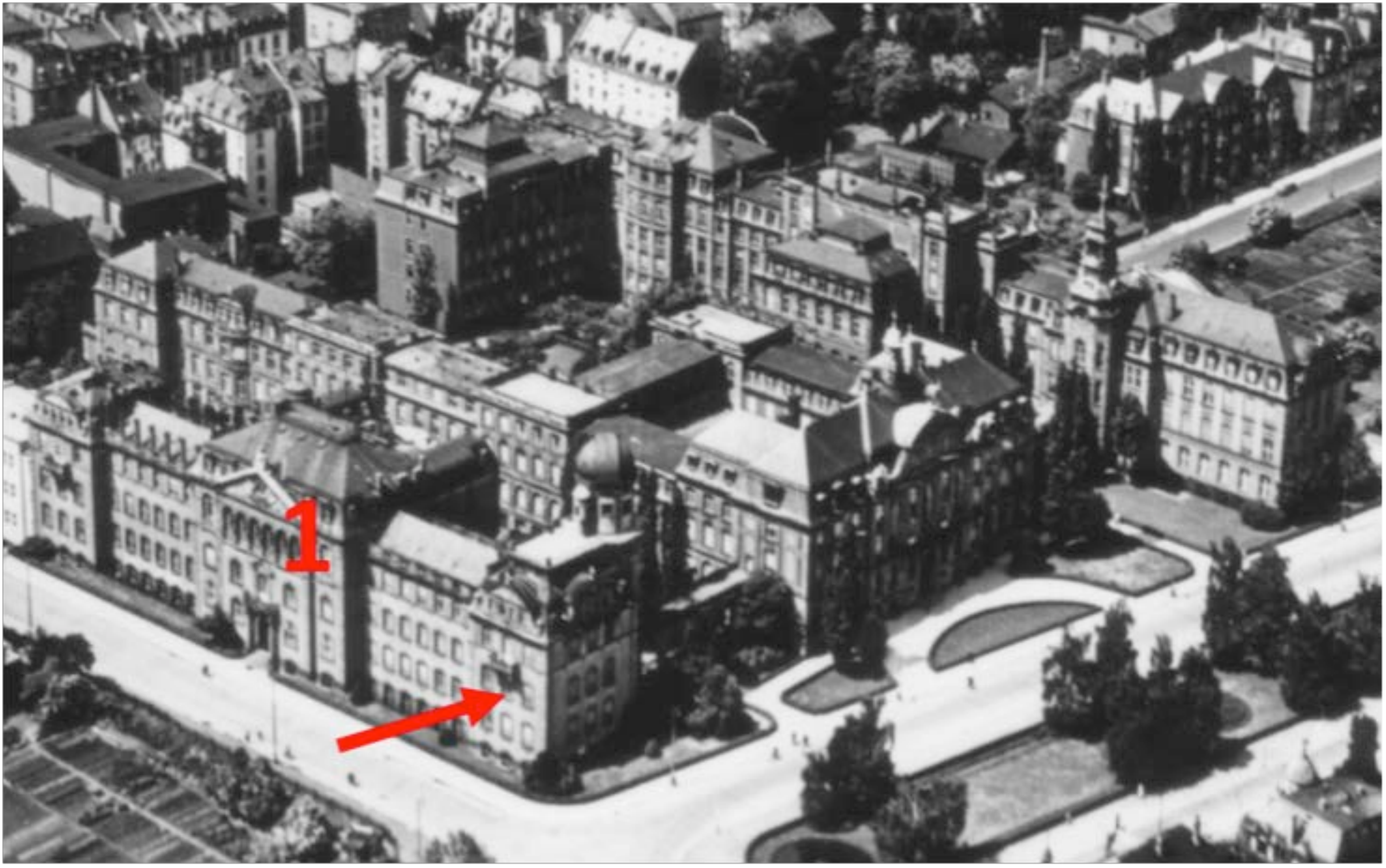}
\caption{Physics building (1) of the University of Frankfurt around 1920. The room, where the Stern-Gerlach experiment was performed, is marked by the arrow (Picture: Gy-UAF).}
\label{fig:physics_building}
\end{center}
\end{figure}
\begin{figure}
\begin{center}
\includegraphics[scale=0.45]{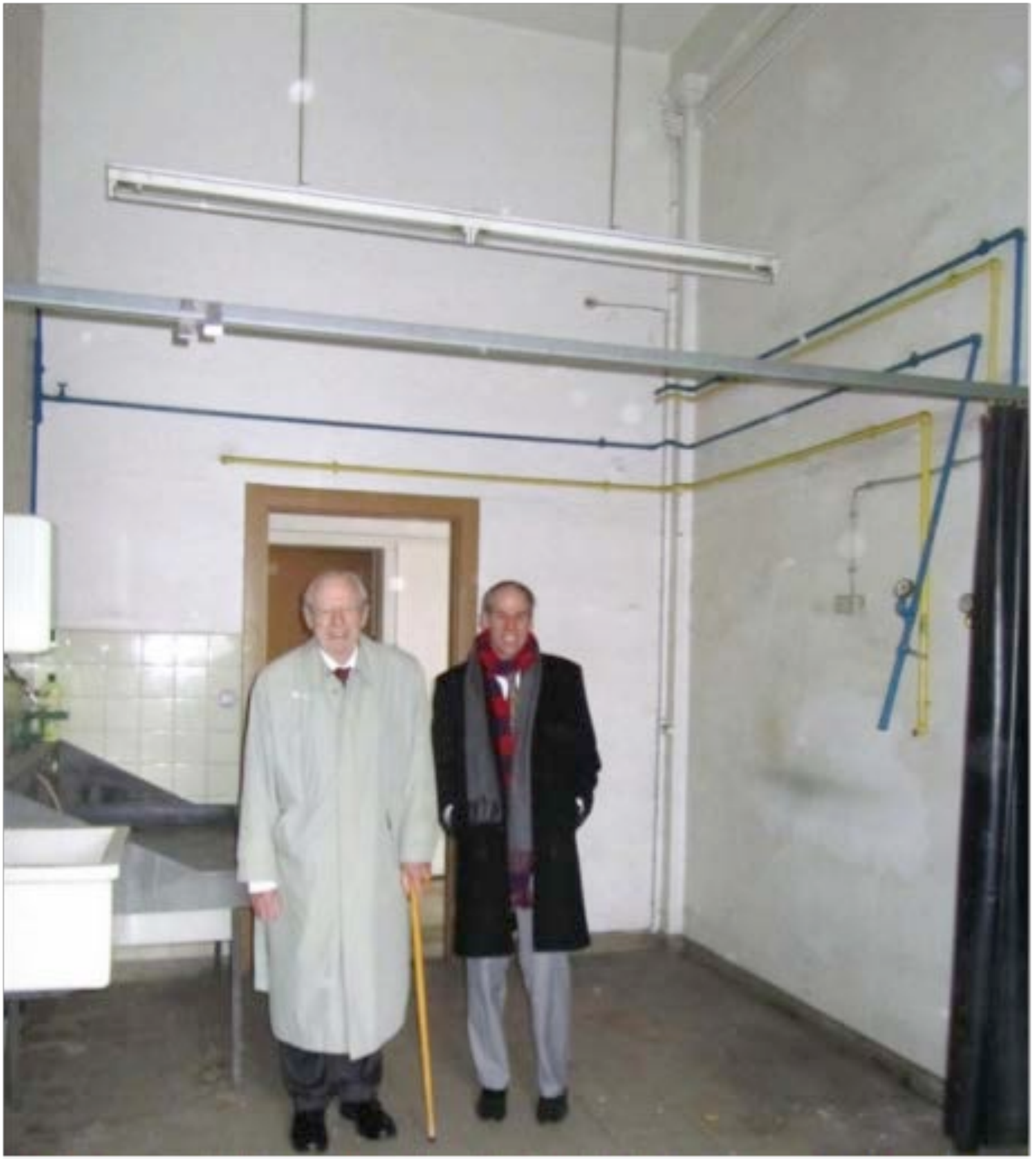}
\caption{The room, where the SGE was performed in 1922. Left: Otfried Madelung, son of Erwin Madelung, who was the director of the theoretical institute in 1922; right: Alan Templeton, grandnephew of Otto Stern. Photo: H.~Schmidt-B\"ocking.}
\label{fig:SGEroom}
\end{center}
\end{figure}

Max Born and James Franck wrote in their nomination letter: 
\begin{quote}
The investigations on magnetic direction quantization by O.~Stern and W.~Gerlach provide us with the most beautiful experimental proof of the existence of discrete quantum states which are here detected by their mechanical properties. In addition, they provide a means to explore the ground states of atoms and the determination of the absolute value of the Bohr magneton. The theoretical foundations of the experiments are due to O.~Stern, but to the actual realization of the very difficult experiments Gerlach was contributing an equal share.\footnote{``Die Arbeiten über magnetische Richtungsquantelung von O. Stern und W. Gerlach bieten uns den schönsten experimentellen Beweis für die Existenz diskreter Quantenzustände, die hier durch ihre mechanischen Eigenschaften nachgewiesen werden. Außerdem liefern sie ein Mittel zur Erforschung der Grundzustände von Atomen und der Bestimmung des Absolutbetrages des Bohrschen Magnetons. Die theoretischen Grundlagen der Versuche stammen von O.~Stern, jedoch ist bei der experimentellen Durchführung der sehr schwierigen Versuche W.~Gerlach in mindestens gleichem Maße beteiligt wie Stern'' (Sw-RSAS, nomination letter for Stern and Gerlach by M.~Born and J.~Franck, {\bf year???})}
\end{quote}  
Werner Heisenberg wrote in 1932: 
\begin{quote}
The immense importance of Stern's experiments derived from the experimental confirmation of directional quantization. When Stern and Gerlach did their experiments, quantum theory had not advanced to such a clarity that one could have predicted the outcome with certainty. The success displayed the discontinuities, that were known until then only for the energy values, also for the magnetic behavior. This gave quantum theory an important experimental support and also provided incentives for its further clarification.\footnote{``Die außerordentliche Bedeutung der Sternschen Versuche lag zunächst im experimentellen Nachweis der Richtungsquantelung. Als Stern und Gerlach ihre Experimente ausführten, war die Quantentheorie nicht bis zu solcher Klarheit fortgeschritten, dass man das Ergebnis des Versuches mit Sicherheit prophezeien konnte. Der Erfolg wies die Diskontiniutäten, die man bisher nur an den Energiewerten kannte, auch im magnetischen Verhalten nach. Damit erhielt die Quantentheorie eine wichtige experimentelle Stütze und Impulse zu weiterer Klärung.''  (Sw-RSAS, nomination letter for Stern and Gerlach by Werner Heisenberg, 1932).}
\end{quote}
During his later tenure in Hamburg (1923--1933), Stern performed more benchmark experiments in quantum physics. Most important among these were the measurements of the magnetic moments of the proton and deuteron \citep{FrischOEtal1933Ablenkunga,FrischOEtal1933Ablenkung,EstermannIEtal1933Moment,EstermannIEtal1933Ablenkung,EstermannIEtal1933Ablenkung2}.  With his molecular beam method he laid the foundations for many other milestone developments in physics and chemistry (e.g. nuclear magnetic resonance, atomic clock, the maser etc.). 

\begin{figure}
\begin{center}
\includegraphics[scale=0.45]{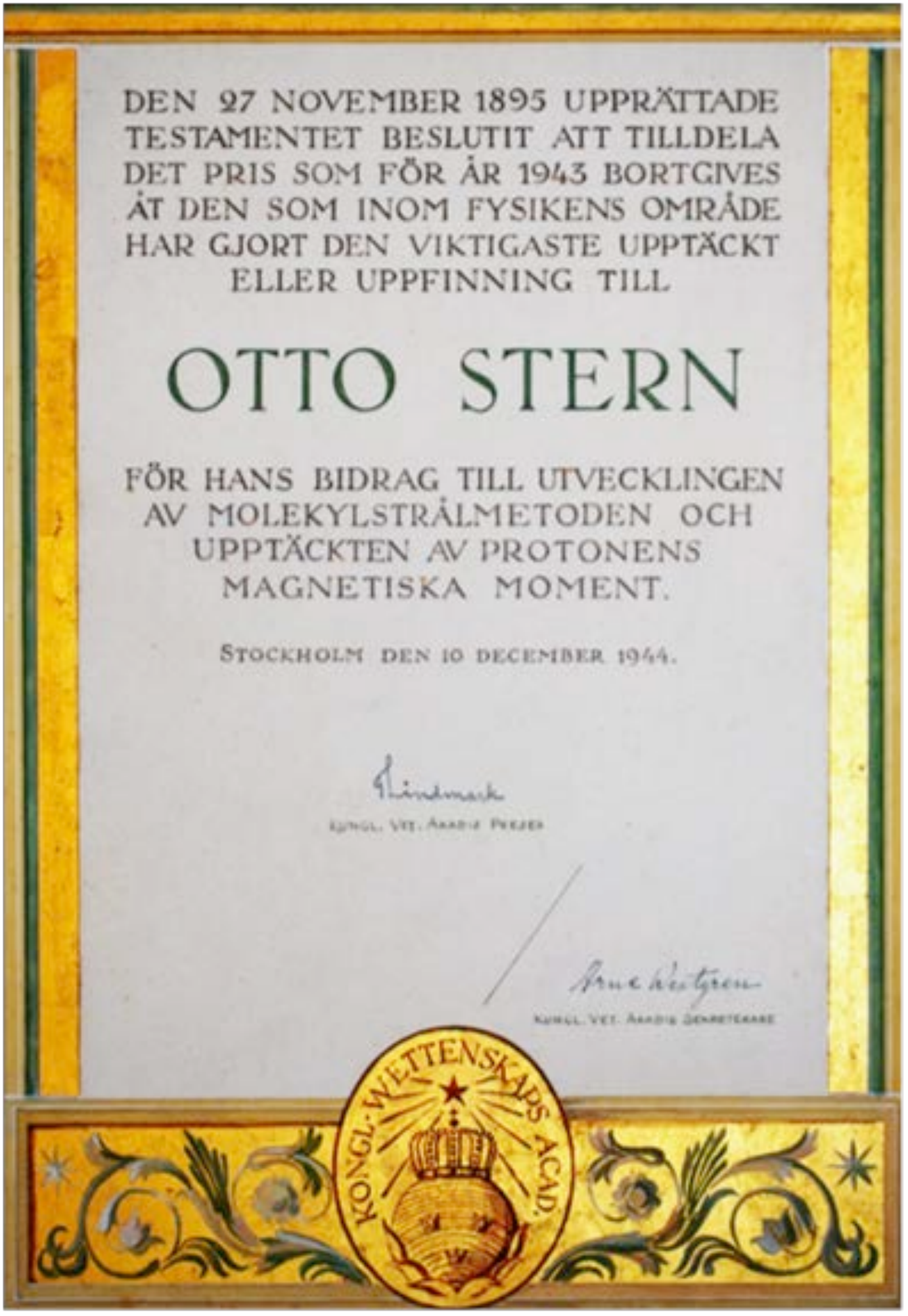}
\caption{Otto Stern's Nobel document (Picture: US-OSF, courtesy Diana Templeton Killen). }
\label{fig:Nobel_doc}
\end{center}
\end{figure}

Stern received the Nobel in physics for the year 1943 for his ``molecular beam method and the measurement of the proton magnetic moment'' (Figure \ref{fig:Nobel_doc}). The decision was made on September 11, 1944 (Sw-RSAS). In the official Nobel award, the SGE is not mentioned but Eric Hulthèn (1894--1981), a member of the Physics Nobel committee, spoke in his prize presentation speech on December 10$^{\rm th}$, 1944, in the Swedish radio mainly about the importance of the SGE. Gerlach who had been the head of the German atomic bomb project (``Bevollmächtigter des Reichsmarschalls für Kernphysik für das deutsche Uranprojekt'') since 1943 was not considered for the prize. Nor was Arnold Sommerfeld, whose hypothesis initiated the SGE, although he was nominated 80 times for the Nobel in physics. 
\begin{figure}
\begin{center}
\includegraphics[scale=0.45]{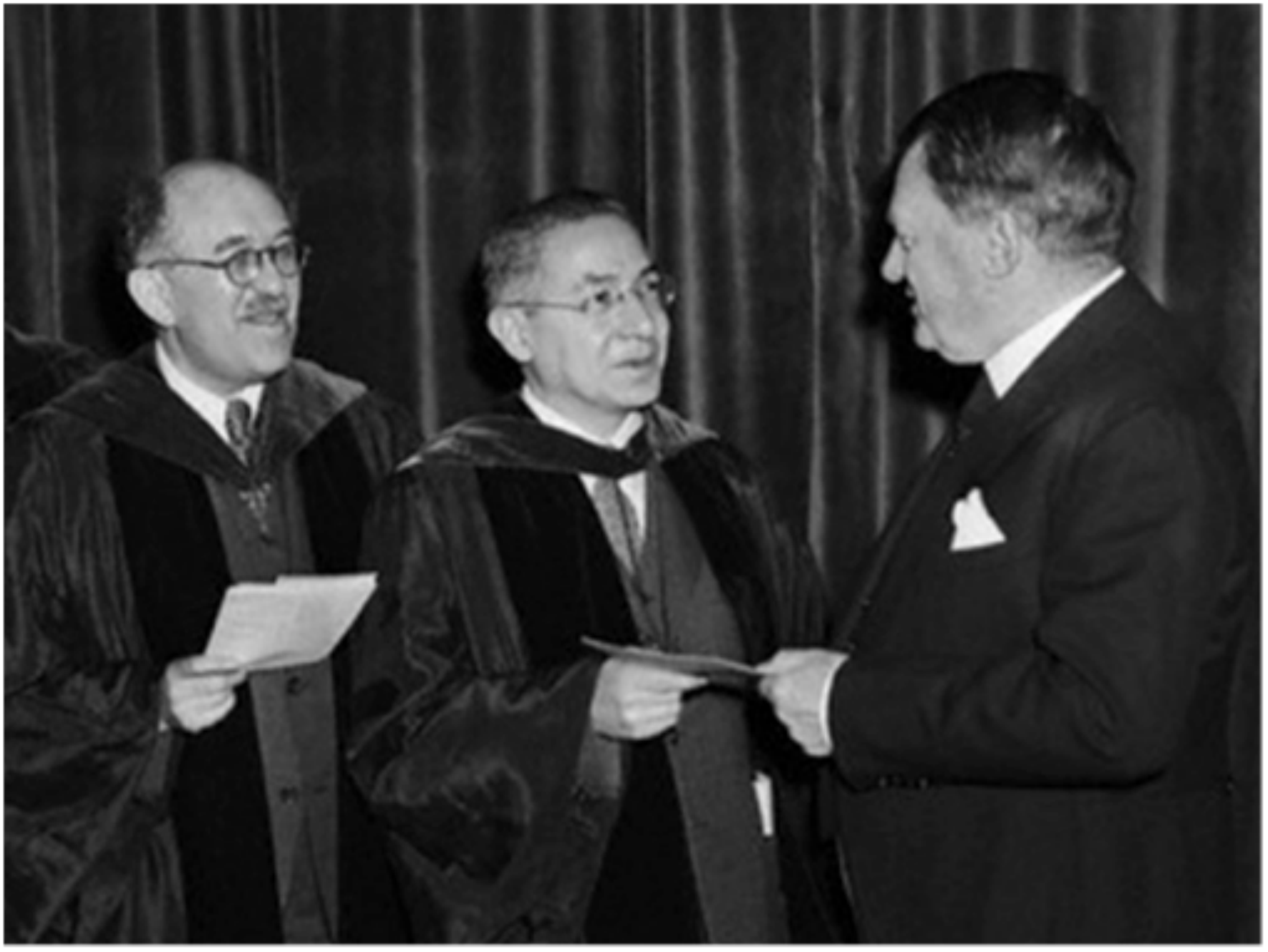}
\caption{The Swedish ambassador Eric Boström presents the Nobel awards in physics to Stern (left) and Rabi (middle) at the New York Walldorf Astoria Hotel on Dec 10th, 1944.  Rabi received the prize for the year 1944 (Sw-RSAS) (Picture: US-OSF). }
\label{fig:Nobel_SternRabi}
\end{center}
\end{figure}

As evidence that the SGE played an important role for awarding Stern the Nobel Prize we quote here from the Presentation Speech of the Nobel Prize in Physics 1943 by the Nobel Committee, held by Erik Hulthèn: 
\begin{quote}
I shall start, then, with a reference to an experiment which for the first time revealed this remarkable so-called directional or space-quantization effect.\\
The experiment was carried out in Frankfurt in 1920 by Otto Stern and Walter Gerlach, and was arranged as follows: In a small electrically heated furnace, was bored a tiny hole, through which the vapor flowed into a high vacuum so as to form thereby an extremely thin beam of vapor. The molecules in this so-called atomic or molecular beam all fly forwards in the same direction without any appreciable collisions with one another, and they were registered by means of a detector, the design of which there is unfortunately no time to describe here. On its way between the furnace and the detector the beam is affected by a non-homogeneous magnetic field, so that the atoms---if they really are magnetic---become unlinked in one direction or another, according to the position which their magnetic axes may assume in relation to the field. The classical conception was that the thin and clear-cut beam would consequently expand into a diffuse beam, but in actual fact the opposite proved to be the case. The two experimenters found that the beam divided up into a number of relatively still sharply defined beams, each corresponding to one of the just mentioned discrete positional directions of the atoms in relation to the field. This confirmed the space-quantization hypothesis. Moreover, the experiment rendered it possible to estimate the magnetic factors of the electron, which proved to be in close accord with the universal magnetic unit, the so-called ``Bohr's magneton''. (Sw-RSAS) 
(www.nobelprize.org/nobel\_prizes/physics/laureates/1943/press.html)
\end{quote}

Otto Stern was forced to leave Germany in September 1933 because he was Jewish. He emigrated to Pittsburgh (USA) and accepted a research professorship at the Carnegie Institute of Technology. On March 8th 1939, he became an American citizen.
He participated in the United States' ``atomic bomb project''. Stern retired at the end of 1945 and moved to Berkeley, where his sister Berta Kamm lived with her family. In 1969, he died in Berkeley of a stroke during a cinema visit. 

Walther Gerlach left Frankfurt at the end of 1924 to become full professor for experimental physics at the University of Tübingen as the successor of Friedrich Paschen. From 1929 on, he was full professor at the University of Munich until his retirement in 1957. From 1943 until the end of the Second World War, Gerlach was directing the ``Fachsparte Physik und die Arbeitsgemeinschaft für Kernphysik im deutschen Reichsforschungsrat'' and from 1944 the German ``uranium project'' (``Uranprojekt''). Gerlach died on August 10, 1979. 

Due to the historic significance of the SGE the old physics building in Frankfurt, where the SGE was performed, was chosen by the European Physical Society in 2014 as an ``Historic Site'' in science, another great honor for Stern and Gerlach. Furthermore, the German Physical Society named its highest award for experimental physicists the ``Stern-Gerlach-Medaille''.

\section{Early attempts to explain the physical processes of directional quantization in the SGE}

As already mentioned above, already a few weeks after the SGE was performed in 1922, Albert Einstein and Paul Ehrenfest tried to explain the SGE results in terms of classical physics \citep{EinsteinAEtAl1922Bemerkungen}. The mechanism responsible for rotating the Ag magnetic moments into directions aligned with the magnetic field remained puzzling. In 1927, Heisenberg  assumed that the process of rotation is made by ``jolting'' (``Schütteln'') \citep{HeisenbergW1927Inhalt}. But this process would need a finite time interval and would require forces to be active. Furthermore, such a ``jolting'' process presumably would continuously try to rotate the magnetic moments during the whole passage through the magnetic field, thus yielding broader and more diffuse spots at the detector. The Schrödinger equation in its standard interpretation, however, implies a sudden ``collapse'' of the magnetic moment at the moment of measurement. 

There has been considerable debate in the literature whether this collapse would, in fact, happen already before the atoms hit the detector screen, in fact, whether it would happen as early as the point of entry when the atom first gets into contact with the $B$-field. Stern himself believed that the silver atom magnetic moment is adiabatically rotated into the observed angle due to Larmor precession \citep{PhippsTEtal1932Einstellung}.  

There is evidence that the SGE was discussed extensively among participants of the 1927 Solvay conference in Brussels. According to notes by participants, the SGE was a topic during the general discussion that took place on Thursday, October 27, 1927. In addition to Bohr and Einstein, participants included H.A.~Lorentz, O.W. Richardson, Paul Ehrenfest, Werner Heisenberg, L. Brillouin, T. de Donder, and others \cite[pp.~436, 478, 500]{BacciagaluppiGEtAl2009Crossroads}. In their discussion, it appears that the issue of a phase loss between the two diverging pathways in the SGE played a certain role, presumably in reaction  to Heisenberg's remarks in his paper on the uncertainty relation \citep{HeisenbergW1927Inhalt}.\footnote{The account in \citep{BacciagaluppiGEtAl2009Crossroads} of the general discussion on the SGE is based mainly on notes that O.W.~Richardson took during the Solvay meeting. Very similar notes of the discussion about the SGE are also extant in the Paul Ehrenfest papers in Leyden (see Ne-LeMB, Ehrenfest Archive, Notebooks, ENB1-32, pages between entries 6609 and 6610).}

\cite{HeisenbergW1927Inhalt} also calculated the statistical distributions within the multiplet states affected by the special process of measurement. He wrote:
\begin{quote}
An atomic beam prepared \`a la Stern-Gerlach is being sent first through a field $F_1$, which is so strongly inhomogeneous that it causes observably many transitions by the action of jolting. Then the atomic beam is running freely for a while, but at a certain distance from $F_1$ a second field $F_2$ begins to act, which is similarly inhomogeneous as $F_1$. It is assumed to be possible that, between $F_1$ and $F_2$ as well as behind $F_2$, the number of atoms in the different states can be measured by means of a possibly applied magnetic field.\footnote{``Ein Stern-Gerlachscher Atomstrahl werde zunächst durch ein Feld $F_1$ geschickt, das so stark inhomogen in der Strahlrichtung ist, da{\ss} es merklich viele Übergänge durch \glqq Schüttelwirkung{\grqq} hervorruft. Dann laufe der Atomstrahl eine Weile frei, in einem bestimmten Abstand von $F_1$ aber beginne ein zweites Feld $F_2$, ähnlich inhomogen wie $F_1$. Zwischen $F_1$ und $F_2$ und hinter $F_2$ sei es möglich, die Anzahl der Atome in den verschiedenen Zuständen durch ein eventuell angelegtes Magnetfeld zu messen.'' \cite[p.~182]{HeisenbergW1927Inhalt}.}
\end{quote}
For a doublet splitting Heisenberg assumed that the wave function amplitudes of the two spin states are in a super-position state and can interfere like the amplitudes of electron scattering on a double-slit. The spin state of each atom is only determined when the atom impacts at the detector. Similar to the analysis of a double slit experiment, Heisenberg predicted that the interference pattern would vary if one would know (i.e.\ measure behind $F_1$) along which path the photon ran or, in the case of the SGE, in which state the atom leaves magnet $F_1$. If one would block off one of the two states behind $F_1$, e.g. by positioning appropriate slits, Heisenberg expected an influence on the distributions behind $F_2$. 

\cite{HeisenbergW1927Inhalt} and Einstein (see below) proposed very similar setups of improved multi-stage SGEs in order to reveal the secrets of the physical processes for directional quantization. To our knowledge, Heisenberg was the first to propose such a multi-stage SGE. \cite{HeisenbergW1927Inhalt} and \cite{GuettingerP1932Verhalten}  (in Wolfgang Pauli's group in Zürich) as well as \cite{MajoranaE1932Atomi} (in Enrico Fermi's group in Rome) made calculations for the probabilities for directional quantization in different multi-stage SGE devices. 

\begin{figure}
\begin{center}
\includegraphics[scale=1.20]{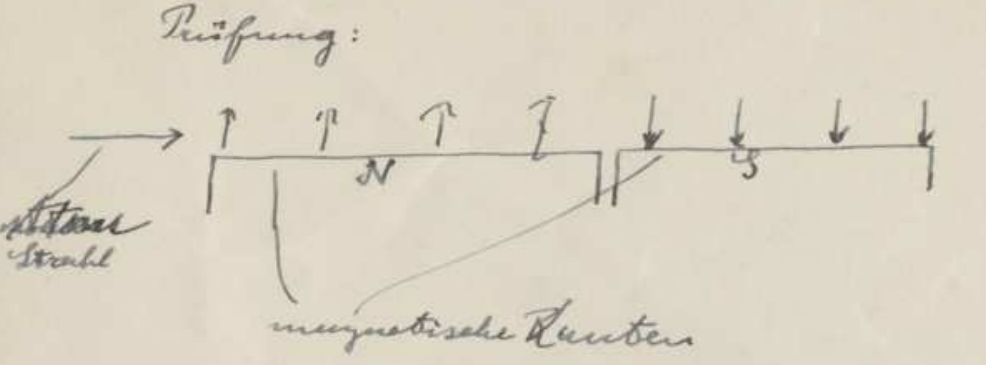}
\includegraphics[scale=0.4]{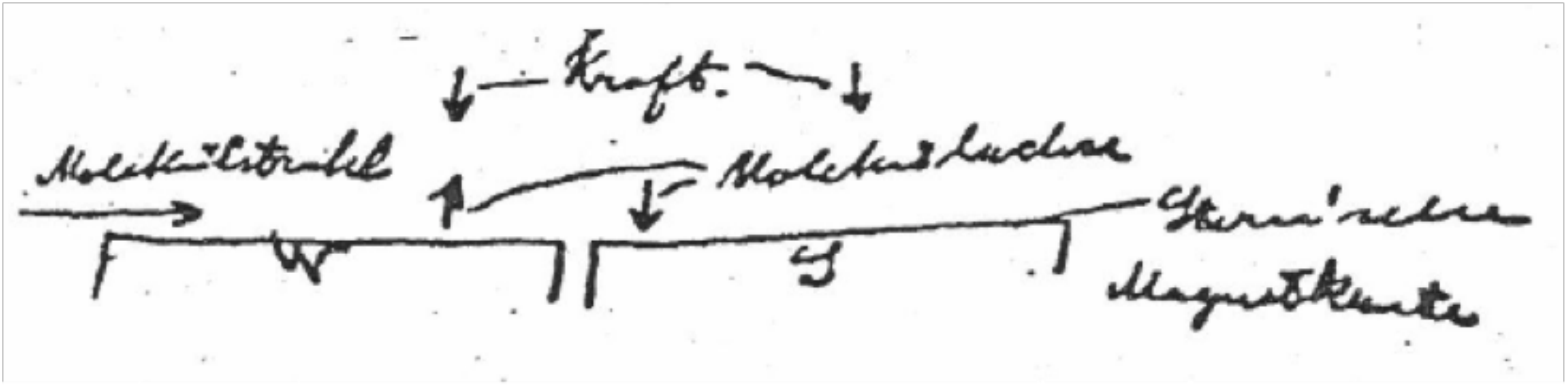}
\caption{Sketches from Einstein's undated letter to Stern (top, US-BL, BANC MSS 85/96 c; Is-AEA~71~120) and to Paul Ehrenfest, dated January 28, 1928 (bottom, Is-AEA~10-173). In the drawings, Einstein indicated the flipping-over of the atomic momenta in a proposed three-stage Stern-Gerlach experiment where an atomic beam entering from left is sent through magnetic fields along the edges (``Kanten'') of two Stern-Gerlach magnets separated by an intermediate region. \copyright{} Albert Einstein Archives, The Hebrew University of Jerusalem, Israel.}
\label{fig:Einstein_sketch}
\end{center}
\end{figure}

In January 1928, a few months after the 1927 Solvay meeting, Einstein wrote a letter to Stern (US-BL, BANC MSS 85/96 c; Is-AEA~71-120) proposing a ``3-stage'' SGE. Similar ideas were communicated at the same time also in a letter to Paul Ehrenfest of 21 January 1928 (Is-AEA~10-173), and the similarity of the ideas, in fact, serves as a basis for the tentative dating of Einstein's letter to Stern to January 1928 (see Figure \ref{fig:Einstein_sketch}).

Einstein hoped that such an experimental device could reveal more information on the process of directional quantization. He addressed Stern ``on the occasion of our quantum seminar'' for help on two  questions in his field of expertise since they ``pertain to the behavior of a molecular beam in a magnetic field.'' It seems that Einstein's letter stimulated Stern to perform an ``improved'' three-stage SGE experiment, where two SG apparatus were combined and aligned to each other. The realization of this experiment by Stern, Phipps, Frisch, and Segrè is described below \citep{PhippsTEtal1932Einstellung,FrischOEtal1933Einstellung}, for a modern discussion of such multi-stage SGE, see, e.g.\ \cite[ch.~XIII]{BohmA1993Mechanics}. 

Einstein's first question was (see also Figure \ref{fig:Einstein_sketch}):
\begin{quote}
I.~An atom aligns itself in a vertical magnetic field this way $\uparrow$ or that way $\downarrow$. The magnetic field is slowly changing its direction. Will each individual atom follow the field in its orientation?\\
Test:\\
Two inhomogeneous magnetic fields, pointing in opposite directions, are being traversed one after the other by the atomic beam. Let one atom be oriented in such a way that it is being deflected upwards in the first field. If it is turned around on transition to the second field, the observed signal would be the same as if both fields would point to the same direction, because of the change of both field and dipole. This is all the more paradox, since the deflection effect grows linearly with the field strength.\footnote{%
``I.~Ein Molekül Atom stelle sich in einem vertikalen Magnet so $\uparrow$ oder so $\downarrow$ ein. Das Magnetfeld ändere langsam seine Richtung. Geht dann jedes individuelle Atom mit dem Felde in seiner Orientierung mit?\\
Prüfung:\\
Zwei entgegengesetzte inhomogene Magnetfelder werden vom Atomstrahl nacheinander durchlaufen. Ein Atom sei so orientiert, dass es im ersten Feld nach oben abgelenkt wird. Dreht es sich beim Übergang zum zweiten Feld um, so muss wegen Umkehr des Feldes und Umkehr des Dipols der Ausschlag genau so ausfallen, wie wenn beide Felder gleich gerichtet wären. Dies ist umso paradoxer, weil ja der Ablenkungseffekt mit der Feldstärke linear anwächst.''
(Einstein to Stern, n.d., US-BL, BANC MSS 85/96 c; Is-AEA~71~120).}
\end{quote}
The letter then continues to discuss a second fundamental question about directional quantization:
\begin{quote}
II.~It is characteristic of our present conception that the field determines the alignment of the atom, its gradient the magnitude of its deflection. The field and its gradient may be varied fully independently. If we imagine that the gradient of the field is fixed and the field varied, then only the direction of the latter is supposed to be relevant, not its magnitude. The field can be arbitrarily weak without effecting the deflections. By a mere change of the direction of an arbitrarily weak field, one may completely change the deflection. This is surely paradox but a necessary consequence of our present conception.\\
Perhaps it might be advisable to produce the inhomogeneous field by means of a water-cooled tube carrying an electric current. If you already are aware of facts that would decide the two questions I would be grateful if you could communicate them to me. If that is not the case, it might be worthwhile to study those questions experimentally.\footnote{%
``II. Für unsere gegenwärtige Auffassung ist charakteristisch, dass das Feld die Einstellung des Molekül Atoms, der Feldgradient die Grösse der Ablenkung bestimmt. Feld und Feldgradienten können ganz unabhängig voneinander var[i]iert werden. Denken wir uns die Feldgradienten gegeben und das Feld var[i]iert; so soll von letzterem nur die Richtung, nicht aber die Grösse massgebend sein. Das Feld kann also beliebig schwach sein, ohne dass dies auf die Ablenkungen Einfluss hat. Durch blosse Änderung der Richtung des beliebig schwachen Feldes sollten also die Ablenkungen völlig geändert werden können. Dies ist gewiss paradox aber bei unserer Auffassung nicht anders zu denken.\\
           Vielleicht wäre es zweckmäßig, das inhomogene Feld durch ein Strom-durchflossenes, wassergekühltes Röhrchen zu erzeugen. Wenn Sie schon Thatsachen haben, welche die beiden Fragen entscheiden, so bitte ich Sie, mir dieselben mitzuteilen. Wenn dies aber nicht der Fall ist, so würde es sich wohl lohnen, diese Fragen experimentel[l] zu bearbeiten.''
(ibid., US-BL, BANC MSS 85/96 c; Is-AEA~71~120).
}
\end{quote}

In Einstein's proposed experiment, the different magnetic momenta are directionally quantized in the magnetic field of the first magnet and dynamically separated when passing through magnetic field 1. In the region between the magnets the selected momenta can be manipulated by varying fields, e.g.\ they can be rotated by a weak magnetic field. In a third stage, consisting of the second magnet, the manipulated momenta are analyzed again. 

Einstein's three-stage SGE is very similar in its basic concept to Isidor Rabi's famous ``nuclear magnetic resonance'' apparatus \citep{RabiIEtal1934Moment,RabiIEtal1939Method,KelloggMEtal1939Moments} which Rabi later designed and used at Columbia University. Rabi recognized that manipulation of the atoms in the region between the magnets can be done by photon excitation (see also \cite[pp.~191--192]{HeisenbergW1927Inhalt}, where he mentioned, that already Bohr in 1927 suggested such a photon excitation process). Because photon absorption is a resonance process as a function of the excitation energy, one may obtain excellent resolution in selecting single energy transitions (e.g. today by very narrow laser lines). This invention made Rabi's apparatus a very high resolution instrument enabling numerous milestone experiments at Columbia University and at MIT. 

When Rabi was working in Stern's group in Hamburg in 1928, he contributed new creative ideas to the further development of the molecular beam technique \citep{RabiI1929Methode}. He showed that an SGE can also be performed in homogeneous magnetic fields. If an atomic beam enters the field in a direction that is not perpendicular (i.e. not under $90^{\rm o}$) to the $B$ field, then dynamical separation of the different spin states is also achieved. This modification of the original SGE setup was a great advantage since homogenous fields can be measured easier and with much higher precision than inhomogeneous fields. 

One might ask: Why did neither Stern nor Rabi ever acknowledge Einstein's letter in any of their papers? Rabi could have known Einstein's proposal, too, because he was a visiting fellow in Stern's laboratory from 1927--1928. Rabi came from Copenhagen (from Niels Bohr) to Hamburg in the fall of 1927 to work with Wolfgang Pauli \citep{SchmidtBoeckingHEtal2011Stern,ToenniesJEtal2011Stern}. There he met Otto Stern and joined Stern's group in doing research with Stern's molecular beam methods. Einstein had written his letter to Stern in January 1928 or before. It seems not unlikely that Stern would have discussed Einsteins proposed experiment with his group members, i.e. also with Rabi. In any case,
Einstein's proposed experimental setup and Rabi's realization are indeed quite similar. 

No response by Stern to Einstein's letter has been found in the Stern papers at Bancroft library (US-BL) or in the Einstein archives (Is-AEA). But Stern began to design such a multi-stage SGE in 1930 \cite[p.~186]{PhippsTEtal1932Einstellung}. In 1931, Stern and Thomas Erwin Phipps, a Guggenheim fellow, who came from the US to work for a few months in Stern's institute in Hamburg, published a paper ``on the alignment of directional quantization''   
\citep{PhippsTEtal1932Einstellung}.
The paper is de-facto an answer to Einstein's letter. The paper does not mention Einstein by name and Stern only indicates in the introduction that the experiment had been proposed ``repeatedly.'' The publication consists of two parts: in a first part, Stern analyzes Einstein's ideas and expresses concerns. In the second part of the publication, Phipps describes the apparatus and reports first measurements.

Stern's concerns are mainly about experimental difficulties. His reaction to Einstein's proposal shows his unique capacity to think  ``out-of-the-box''. The proposal inspired him to design an ingeniously simple apparatus.  His concerns were the following:
\begin{quote}
The question now is how directional quantization in the second field comes about. In the course of the development of the new quantum mechanics, this problem has been treated repeatedly in a theoretical way. The result was that part of the atoms align parallel, part anti-parallel, e.g. in the case of a right angle between the two field directions, one half parallel, the other half anti-parallel.

But it always seemed certain to me that if one would actually carry out the experiment nothing of the kind should be expected. Instead all atoms would rather follow the rotation of the field without “flipping over”. This is because in the above-mentioned calculations it was always tacitly assumed that the change of the field direction would take place strictly non-adiabatically, But in reality the contrary is the case under the experimentally attainable conditions. The rotation of the field direction has to be regarded with great approximation fully as an adiabatic process, because the atom carries out a great number of Larmor rotations while it traverses a path over which the field direction changes.\footnote{``Die Frage ist nun, wie sich in dem zweiten Feld die Richtungsquantelung einstellt. Nach Entwicklung der neuen Quantenmechanik ist dieses Problem mehrfach theoretisch behandelt worden mit dem Resultat, daß sich ein Teil der Atome parallel, ein Teil antiparallel einstellt, z.B. für den Fall eines rechten Winkels der beiden Feldrichtungen die eine Hälfte parallel, die andere Hälfte antiparallel.\\
Es schien mir jedoch seit jeher sicher, daß bei wirklicher Ausführung dieses Versuches nichts derartiges zu erwarten wäre, sondern alle Atome der Drehung des Feldes folgen würden, ohne ``umzuklappen''. Denn bei den erwähnten Rechnungen war stets stillschweigend die Voraussetzung gemacht worden, daß die Änderung der Feldrichtung streng nicht-adiabatisch erfolgt. In Wirklichkeit ist aber unter den experimentell herstellbaren Bedingungen gerade das Gegenteil der Fall, die Drehung der Feldrichtung muß mit großer Näherung als durchaus adiabatischer Prozeß betrachtet werden, weil das Atom eine große Anzahl Larmordrehungen ausführt, während es eine Strecke durchfliegt, auf der sich die Feldrichtung ändert.'' \cite[pp.~185--186]{PhippsTEtal1932Einstellung}}
\end{quote}

For the theoretical calculations mentioned in the quote, papers by Charles Galton Darwin (1887--1962) and Alfred Land\'e (1888--1976) were cited \citep{DarwinC1928Motion,LandeA1929Polarisation}.

As discussed above, Einstein and Ehrenfest had estimated the alignment of the silver magnetic moments by a process of classical Larmor radiation.
Similarly, Stern was convinced that the atomic magnetic moment undergoes a Larmor precession (continuous in angle), and he believed that this Larmor precession is turned by an unknown process adiabatically into the final orientation.  As shown by \cite{HermansphanNEtal2000Observation}, this assumption is a purely classical one and contradicts quantum dynamics.

\begin{figure}
\begin{center}
\includegraphics[scale=0.4]{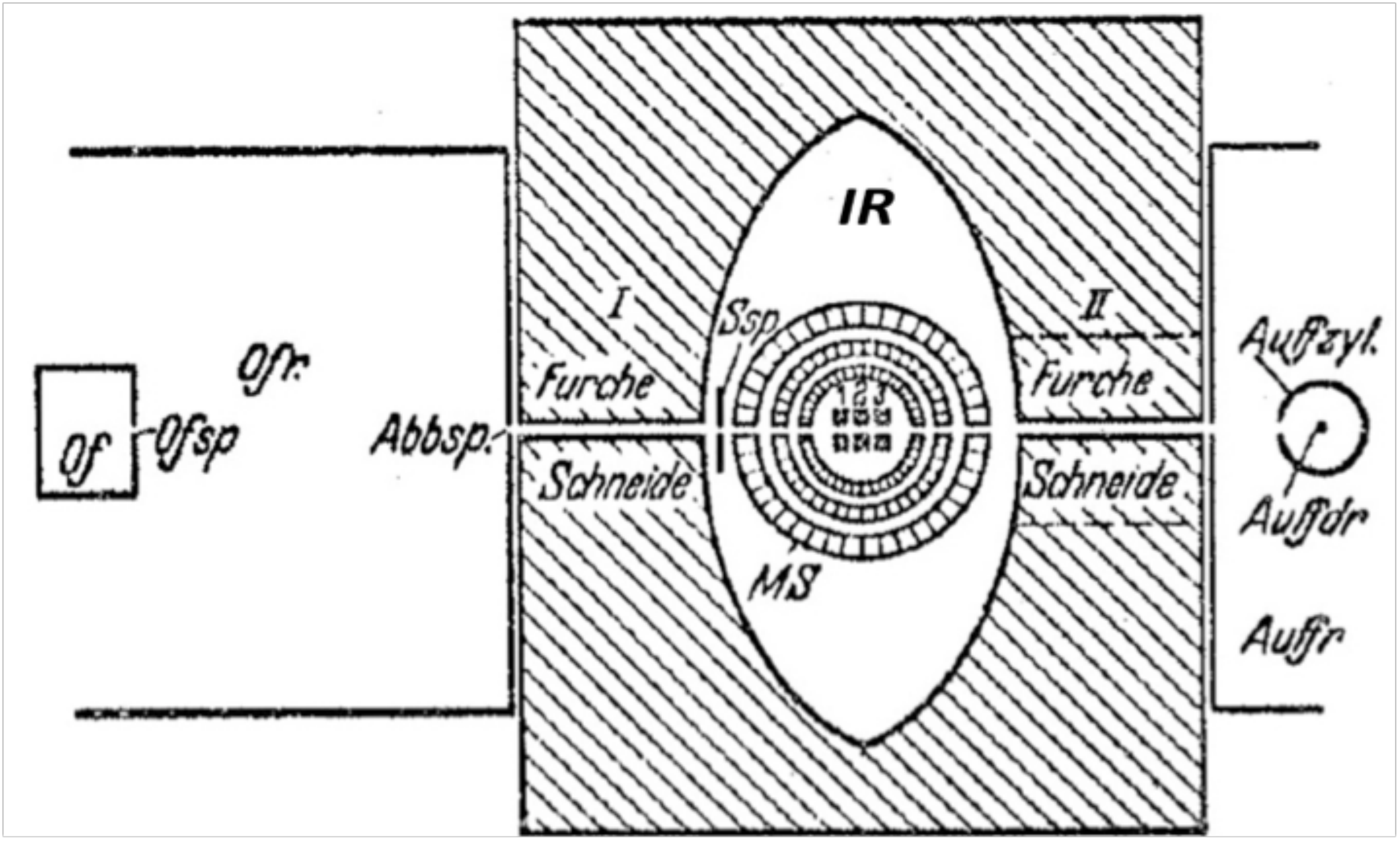}
\caption{Experimental setup of the Phipps-Stern experiment: Of Oven, Ofsp Oven aperture, Ofr oven region, I and II first and second magnetic field, Ssp Selection slit, MS magnetic shielding, 1,2,3 small electro magnets, IR inner region, Auffzyl Selector cup, Auffdr detector wire, Auffr detector area \cite[p.~189]{PhippsTEtal1932Einstellung}.}
\label{fig:PhippsSternSetup}
\end{center}
\end{figure}

The apparatus of \cite{PhippsTEtal1932Einstellung} had three separated field regions combined together (magnet $M_1$, inner region IR with varying fields, and magnet $M_2$, the whole apparatus was about $20$~cm long, see Figure \ref{fig:PhippsSternSetup}). Behind the field region of $M_1$ one could separate off one fraction of the atomic beam by setting the slit in such a way that only part of the beam can get pass through it (Ssp), Thus the magnetic momenta of this fraction point only in one direction. This selected fraction passed through region IR, which was made nearly field-free by a very good magnetic shielding. Thus the field inside IR was about a factor $1000$ weaker than in magnet $M_1$. 

Inside region IR three very small electro magnets were mounted. Their fields were rotated relatively to each other by $120^{\rm o}$. With the help of these three small magnets the field in region IR could be varied linearly in the range of a few Gauss. Thus the moving atom experienced a time-varying field. Because the time $T_L$ of one Larmor rotation of the atomic magnets would be very short compared with the time of flight $T_F$, which the atoms needed to pass the region along the small magnets, as Stern expected, the direction of the atomic momenta of the atoms should adiabatically be rotated and follow its rotation created by the three small magnets. Thus, if this rotation can be varied between $0^{\rm o}$ and $360^{\rm o}$ the atomic momenta of the atomic beam, after leaving region IR, could be steered in different directions. This beam consisting of atoms with momenta rotated in this manner is now sent through magnet $M_2$ (identical to magnet $M_1$ with parallel field direction). In $M_2$ the beam which was manipulated in the IR region is analyzed with respect to the directions of its magnetic moments. 

Stern calculated the path length for a moving atom per one Larmor rotation as a function of the field strength in region IR. Only when the fields inside region IR can be made as small as one Gauss or so the correct adiabatic conditions could be obtained. Stern contacted Paul Güttinger (1908--1955) of Wolfgang Pauli's group in Zürich to calculate the probabilities for flipping-over of the atomic momenta in the above described field configurations on the basis of the new quantum theory \citep{GuettingerP1932Verhalten}. Güttinger predicted a flipping-over probability of 6\% for a field strength of one Gauss and a path-length of one mm.
\begin{figure}
\begin{center}
\includegraphics[scale=0.4]{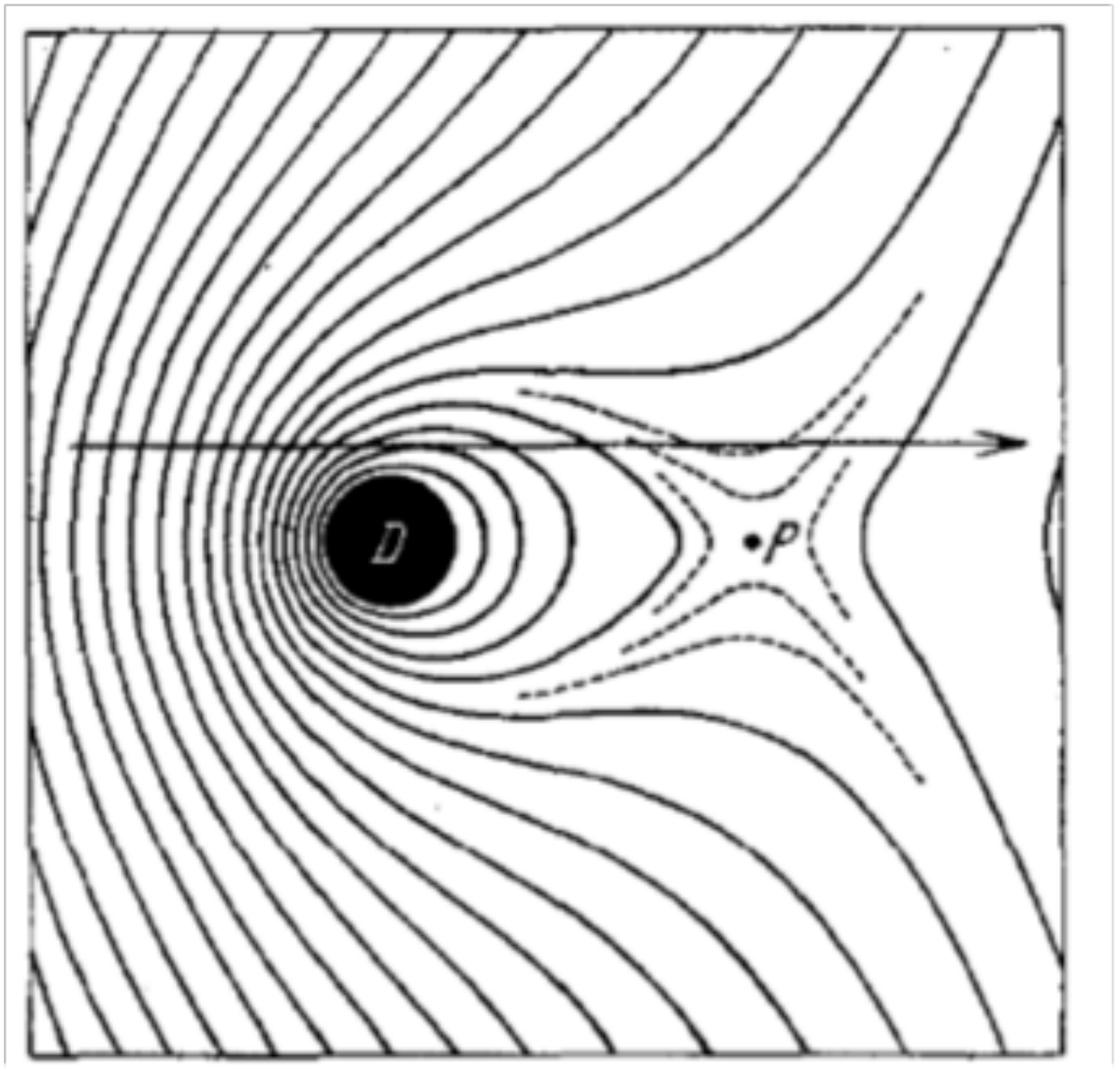}
\caption{Wire $D$ with magnetic field lines induced by the electric current $i$. The arrow indicates the atomic beam path and $P$ is a point where the field vanishes \cite[p.~611]{FrischOEtal1933Einstellung}.}
\label{fig:FrischSegreField}
\end{center}
\end{figure}

Although Phipps obtained very good position resolution of the beam spot behind magnet $M_2$, using a very thin wire (see Figure \ref{fig:PhippsSternSetup}: Auffdr) as a detector (i.e.\ using the so-called Langmuir-Taylor-Method \citep{LangmuirI1925Effects}), the results he obtained were not consistent. He did not observe any ``flipping-over'' of atomic magnetic momenta at all. Only in a single measurement did he find some portion of flipped-over magnetic momenta. But in this particular measurement run the vacuum in the inner region IR had been damaged because a wire insolation of one of the small magnets had been burnt. Therefore Stern concluded that this particular observed instance of a flipping over was induced by scattering of the sodium atoms at rest gas molecules rather than induced by the applied magnetic field. Unfortunately, Phipps could not continue these experiments, since he was forced to go back to the USA.

Robert Otto Frisch and Emilio Segrè repeated Phipps experiment with a new setup under improved conditions \citep{FrischOEtal1933Einstellung}, \cite[pp.~68--71]{SegreE1993Mind}.  Most importantly, instead of the three small magnets in region IR they mounted a wire, probably following Einstein's advice. In a demonstration of his remarkable experimental skills, Einstein had proposed a clever technical trick for producing controlled magnetic fields to rotate the magnetic momenta in the region IR.  His idea was to mount just a wire $D$ (``stromdurchflossenes Röhrchen'') inside region IR and vary its current. Sending a current $i$ through the wire, the required magnetic field in region IR could be created, ensuring controlled adiabatic conditions. The new inner region is shown in Figure \ref{fig:FrischSegreField} together with the magnetic field lines. The beam path is indicated by the arrow. 
At point $P$ the magnetic field vanishes. They also changed the design of the experimental setup a little bit by moving the location of the selection slit, and they used potassium atoms instead of sodium.

\begin{figure}
\begin{center}
\includegraphics[scale=0.4]{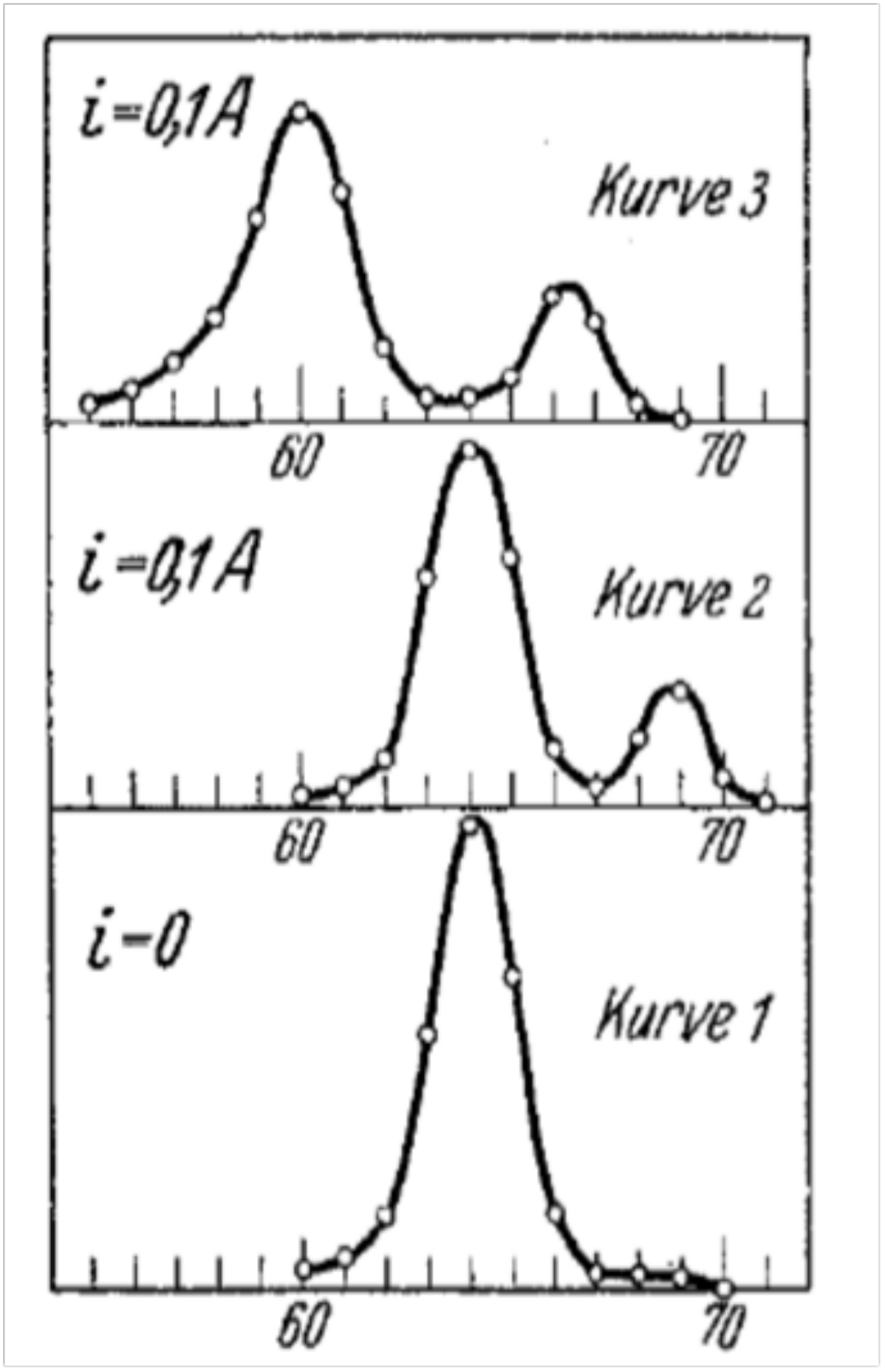}
\caption{Beam intensity distributions behind magnet $M_2$; the small peak of curve 2 and 3 originates from flipped-over atoms. $i$ is the electric current in wire D \cite[p.~614]{FrischOEtal1933Einstellung}.}
\label{fig:FrischSegreResults}
\end{center}
\end{figure}

Since Güttinger's calculations had been based on the old field configuration of the Phipps design, Frisch and Segrè needed calculations for the new field configuration. For these calculations they could refer to a paper by Ettore Majorana \citep{MajoranaE1932Atomi,InguscioM2006Comment}. For the flipping probabilities $P$, Majorana obtained the formula $P=e^{-k\pi/2}$, where $k$ is a function of the current $i$, the velocity $v$ of the atomic beam, and the path length (or flight time $T_F$) inside IR in the neighborhood of $P$. Majorana extended his calculations for a non-vanishing field inside IR, when $i$ was zero. Figure \ref{fig:FrischSegreResults} shows the experimental results of Frisch and Segrè. 

In Figure \ref{fig:FrischSegreResults} , ``Kurve 1'' shows the intensity distribution for $i=0$. Obviously no, or very little, rotation of the magnetic moments occurred in region IR (there may be a tiny shoulder at the abscissa value of $68$, which is the position, where flipped-over momenta would be observed). ``Kurve 2'' is obtained for an electric current of $0.1$~Ampere. About 25\% of the beam underwent flipping-over. ``Kurve 3'' was obtained under the same field conditions as ``Kurve 2'' but here the slit (Ssp) was slightly shifted. The comparison of curve 2 and 3 shows that the atom beam trajectories are classical and well controlled. The flipping-over probability and the distance between the two peaks is approximately the same for both curves. These experiments by Phipps, Frisch, and Segrè showed that one can quantitatively describe the classical trajectories of the beam fractions in the whole three-stage system, but not the flipping-over probabilities. 

To explore the dynamics of the passage of the sodium or potassium atoms through the three-stage system one might argue that each atom that passes through the magnetic field regions together with the entire SGE apparatus represents a closed system, where dynamical conservation laws of momentum and angular momentum strictly apply at any moment. 

\cite{WennerstroemHEtal2012experiment,WennerstroemHEtal2013measurements,WennerstroemHEtal2014Interpretation} recently calculated for a single atom passage the trajectory and the directional quantization probabilities in the original SGE. 
They assume the magnetic moment of each atom to be stochastically coupled to the spins of the atoms in the magnet. Using different transition times for this coupling they obtained the position distribution probability on the detector behind the magnet. They found good agreement with the results of Gerlach and Stern when either the transition time was very fast compared to the whole passage time or the transition occurred like a sudden collapse at the entrance into the magnet.  

A stochastic coupling process of the Ag magnetic moment with the magnet spins could probably occur, however, at any position of the passage inside the magnet again and again for Ag atoms in both directions. Thus, in case of a stochastic process one would expect a more diffuse beam spot on the detector behind the magnets which is, however, experimentally not found. Therefore, along this reasoning, the spin orientation should already have been decided at the entrance of magnet $M_1$ .

According to experimental observations the probability for the magnetic moment of each atom to align with respect to the magnetic field of magnet $M_1$ is $100\%$ independent of the details of the field configuration before. If we have a fully random distribution in the angular orientation before entering magnet $M_1$ then we expect $50\%$ of the beam intensity in both spin orientations. If the distribution is not random, e.g.\ if it is being manipulated in region IR, then the ratio of the two doublet peaks after magnet $M_2$ can vary. Frisch and Segrè did however, not find satisfactory agreement between the Majorana theory and experiment. Majorana predicted an increase of $P$ up to $100\%$ for increasing current $i$. Frisch and Segrè observed a maximum of $25\%$ at about $i=0.1$~Ampere and then they saw a decrease at higher currents. The fraction of flipped-over magnetic momenta in region IR apparently depends on fine details of the field in region IR. These details of the apparatus inside IR were not known or controlled precisely enough in order to get a trustful agreement between experiment and theory.

The experiment of Frisch and Segrè allows the following conclusions: 
\begin{enumerate}
\item The momentum orientation selected by means of $M_1$ remains with $100\%$ probability in the same oriented state after passing magnet $M_2$, if the region IR is field-free and both magnetic fields $M_1$ and $M_2$ are parallel. This observation confirms the projection postulate. If, after passing magnet $M_2$, the atoms continue to move in field- or collision-free areas the orientation of the atomic magnetic moments will remain unchanged forever. 
\item The probabilities $P$ for finding up- or down-oriented momenta behind a magnet depend on the angular momentum orientation before the atoms enter the magnetic field (called initial-state distribution). 
\item If the field direction of magnet $M_2$ would be rotated relative to the field direction of magnet $M_1$ the detected doublet structure behind magnet $M_2$ would be aligned only with respect to the field direction of magnet $M_2$. 
\item The sharp line structures on the detectors suggest that the directional quantization occurs at the entrance of the magnet and not inside the magnetic fields of magnets $M_1$ or $M_2$. 
\end{enumerate}

\section{How to describe the path of a single atom through the SG apparatus: classical trajectory or superposition of wave packets?}

Does the Frisch-Segrè-Experiment provide any answer to the question whether super-position of the two $\pm$-spin states has an influence on the results of the experiment? By setting apertures behind $M_1$, Frisch and Segrè selected a single spin state and injected it into the field of magnet $M_2$ (having the same field direction). The atom injected into $M_2$ therefore carries information about its spin orientation. If superposition occurs in the field of magnet $M_2$ each single atom should be in a super-positioned state of the wave packets of both spin states ($\pm$). Both states should interfere and always produce some double peak structures on the detector. The experiment, however, clearly showed that only the one selected spin state was observed behind $M_2$. Thus, it strongly suggests that in an SG-like experiment single atoms move on classical trajectories which can be determined by measuring the transverse momentum of each atom.

\cite{DevereuxM2015Reduction} and \cite{WennerstroemHEtal2012experiment,WennerstroemHEtal2013measurements,WennerstroemHEtal2014Interpretation} discuss the long and continuing debate how to describe the path of the atoms through the SGE apparatus: is the path created by super-position of the wave packets of the two spin orientations or can one view the path as a classical trajectory? 
As the philosopher Karl Popper (1902--1994) already recognized, it makes a difference whether one views quantum measurements in direction of ongoing time as a ``predictive measurement'' (experiment) or as a ``retrodictive'' measurement after detection of the atom \cite[sec.~77, App.~vi]{PopperK1989Logic}, \cite[pp.~60--64]{PopperK1982Theory}. In a predictive experiment, the initial spin orientation is in general unknown. Applying the laws of quantum dynamics, the wave packets of the two spin states are in a super-position state and might interfere. The question remains as to how long two wave packets are in a coherent state on their path through the SG magnet and whether an interference would be observable? 

The SGE is often compared with the scattering of photons or electrons on a double-slit system. In such a double-slit scattering experiment, however, there is no experimental way to identify from the detected photon or electron, either predictively or retrodictively, the path of the projectile, i.e. there is no experimental way to determine through which slit each projectile passed because the slit distance is smaller than the de Broglie wave length of the projectile. The photon or electron carries no measurable fingerprint in which slit it was scattered. Therefore, the detected scattering distribution as a function of the transverse momenta must be calculated from a super-position of the wave packets of the two non-distinguishable pathways, yielding the well-known interference pattern. In his famous discussions with Bohr, Einstein argued if one could measure the tiny transverse momentum given to the slit system after single projectile scattering one could measure through which slit it passed and the interference would disappear \citep{BohrN1949Discussion}.  

In the SGE, on the other hand, one can clearly identify, in the retrodictive case, the spin orientation of each atom. The process of momentum exchange of each single atom with the apparatus proceeds in a very different manner in the SGE  than in the case of photon or electron scattering in a double slit system. When the atoms enter the first ``very thin layer'' of the magnetic field (at the entrance) the directional quantization of the two spin states must occur extremely sudden, but without any measurable amount of momentum exchange similar to the photon or electron case. When the spin oriented atoms are then passing through the SG magnetic field each atom is continuously accelerated  in $z$-direction due to the acting force $\mu\frac{\partial B}{\partial z(t)}$. 

The final transverse momentum in $z$ direction $\pm p_{z0}$  of each spin state after leaving the magnet is $\pm p_z=\int \mu\frac{\partial B}{\partial z(t)} dt$, where the limits of the integral are $t = 0$  and $t = T_F$. This maximum value (which has always been found in all the numerous SGE experiments performed so far) can only be reached if the atomic magnetic moment $\mu$ always points in the $+$ or $-z$-direction during the entire passage through the magnetic field. The final transverse momentum in $z$-direction, measured by the detector,  is a direct measure of the time-integral over the acting force. The SG apparatus measures only this transverse momentum and never spin orientations. 

Indeed, the SGE provides an extremely high-resolution measurement of the transverse momentum.  For information: Each Ag atom in the historic SGE had a mean momentum of $p_x = 54$~a.u.. Since the mean deflection in $z$ direction in the $3.5$~cm long magnet was $0.1$~mm the Ag transverse momentum was $p_z = 0.15$~a.u.. This corresponds to a transverse kinetic energy of $1.6\cdot10^{-3}$~eV. Using narrower slits (e.g.\ $1.5$~micron) and a velocity selected atomic beam the momentum resolution could be strongly improved below $0.005$~a.u. corresponding to an energy change in the atom kinematics of below $10^{-5}$~eV. This excellent resolution value demonstrates the great power of momentum measurements and the great breakthrough of the historic SGE in resolution. 

\cite{HeisenbergW1927Inhalt} discussed the SGE as a benchmark example of the resolution limits in a quantum physics measurement in his famous paper on the uncertainty relation. But he considered only the energy and not the momentum measurement of the magnetic atom in the magnetic field. He concluded that the precision of the energy measurement depends on the time of flight through the magnetic field $t_F$. The precision of momentum measurement in a single scattering event is not restricted by Heisenberg's uncertainty relation.  The uncertainty relation does not affect Stern's measurement approach. Notice, though, that a direct energy measurement would never yield such a resolution.

This precisely measurable exchange of transverse momentum via the magnetic field interaction implies a reduction to a single eigenfunction \citep{DevereuxM2015Reduction}. According to \cite[ch.~21]{BohmD1951Quantum} one can calculate the path for a super-positioned two-spin state wave function and the super position remains until impact on the detector. But in momentum space (because of the extremely small de Broglie wave length) the two spin states after entering the magnetic field region are immediately distinguishable by their transverse momenta and follow detectable classical trajectories. 
This is in full agreement with Bohm's interpretation of the SGE, where he stated: ``It would be possible in principle to measure the spin by measuring the momentum transmitted to the particle by the magnetic field'' \cite[p.~596]{BohmD1951Quantum}.

In recent work of \cite{DevereuxM2015Reduction} and \cite{WennerstroemHEtal2012experiment,WennerstroemHEtal2013measurements,WennerstroemHEtal2014Interpretation} the basic question of super position in the SGE is discussed in detail. Devereux points out: 
\begin{quote}
Observation of two separated beam spots at a detection screen downstream of a Stern-Gerlach magnet does not, in fact, demonstrate that the wavefunction of a neutral spin one-half particle has remained in a spin superposition while traveling through that magnetic field. The wave function may have been reduced to just one spin-direction eigenfunction, as David Bohm suggested, by immediate momentum and energy transfer with the magnet, rather than by subsequent, which-way determination at the screen. 
\end{quote}
And he argues:
\begin{quote}
Thus, following energy exchange with the S.-G. magnet, transferring field quanta, there must be just a single wavepacket representing the spinning particle.  A computation for development of that single wavepacket within the magnetic field agrees with the observations of Stern and Gerlach. And, no empirical evidence substantiates continuation of a spin superposition through the S.-G. magnet, though there are several experiments which deny that explanation. 
\end{quote}
Notice the following. As pointed out above, due to the magnetic force there is continuous momentum exchange between atom and magnet. This transverse momentum exchange enforces a very tiny energy exchange too.

A comparison of the SGE with the detection concept of Aston's mass spectrometer supports this argumentation. Before 1918, dynamical detection approaches in vacuum were never used which is why the SGE represents a milestone invention in detection technique. In 1919, \cite{DempsterAJ1918Method} and \cite{AstonF1919Spectrograph} used a similar kinematical method like Stern would do a little later, to separate an ion beam in vacuum. But they were by far not achieving the momentum resolution as it was obtained by Stern and Gerlach. When one injects a two-component beam of singly and doubly charged He ions in a well-defined momentum state into a magnetic field of Aston's mass spectrometer, their trajectories in a magnetic field are uniquely determined by classical dynamics. The two different trajectories result from the different strength of force acting on the singly and doubly charged ions. Nobody would argue that the wave function of the two He ionic states can interfere and that the charge state wave function would collapse and the ionic charge would be decided only when they hit the detector. If there were a superposition of wave functions of both ion states the beam emittance would be strongly affected. Nobody has ever seen such an effect. 

If there were super-position along the whole path, the distribution of the silver atoms on the detector (see Figure \ref{fig:SGEobservedpattern} (above) of Stern's own slide of the historic SGE) should show symmetries with respect to the magnetic field direction. In the direction of the magnetic field (we consider only the center region of the magnet) one observes the splitting but this is not symmetric for both spin states. The component passing in the middle near the sharp edge of the magnet sensed a higher force and got more deflected (cusp shape) which is a purely classical effect. In case of superposition of both spin states, the cusp should be also present in the other spin state distribution which is not the case. Notice also that the intensity for both spin states is not equally distributed. On the side of stronger field in the middle we observe more deflected atoms. In principle, this asymmetry could be modeled as well with wave packets of a coherent spin superposition that would be deformed and distorted by its passage through an asymmetric field configuration and only collapse at the point of detection at the screen. But the precise image of the asymmetries of the magnet's edges and geometric configuration rather suggests a classical trajectory of atoms with aligned magnetic moments all the way through the SG magnetic fields.

In a ``retrodictive measurement'' even in a multi-stage SGE one can achieve complete information on the path of an atom through several magnetic fields by the transverse momentum measurement. From the path determined in this way the orientation of the spin states can be deduced. To illustrate the detection power of high resolution momentum spectroscopy in a SG apparatus the momentum measurement in a  ``two stage'' SGE will be discussed here. 

\begin{figure}
\begin{center}
\includegraphics[scale=0.4]{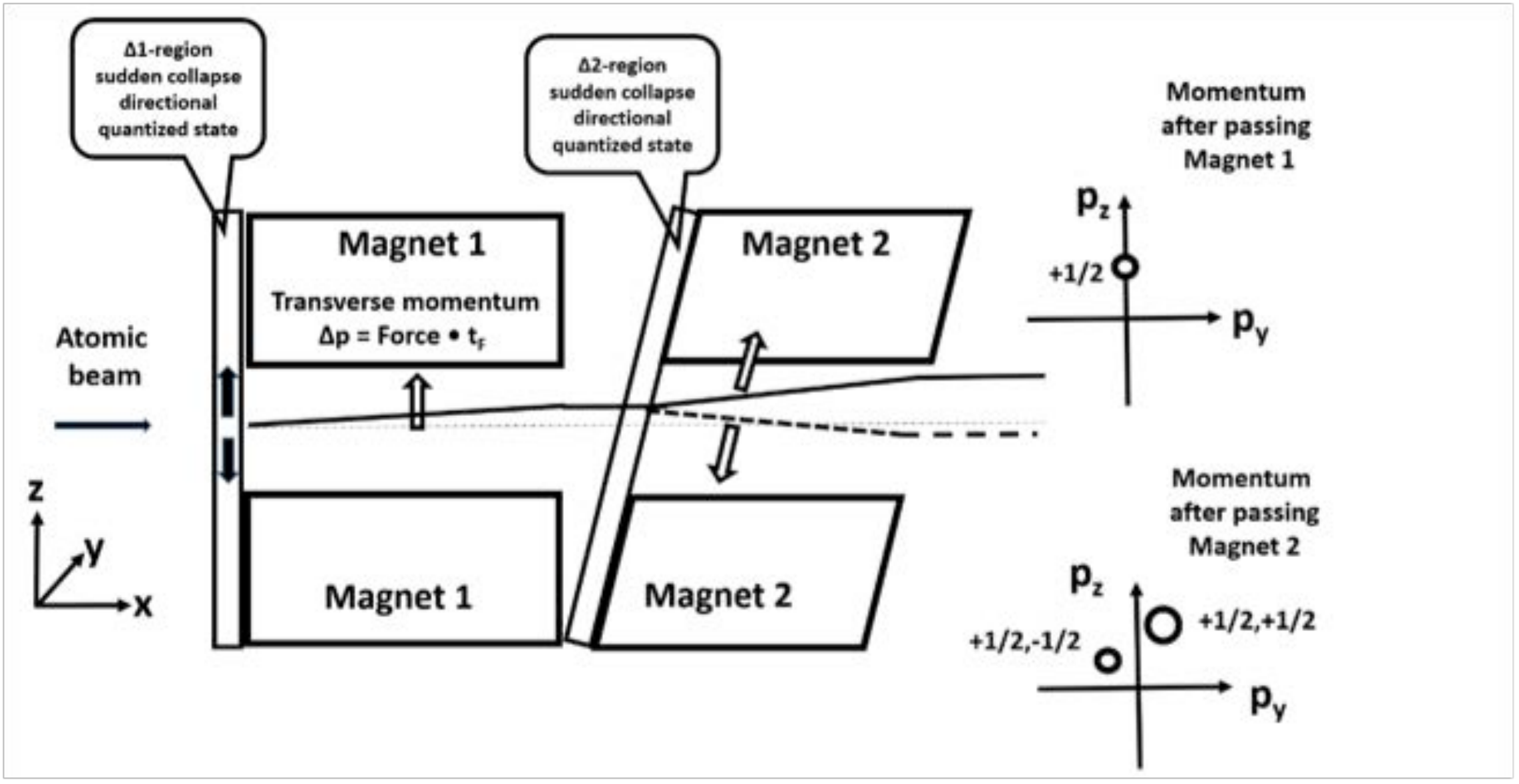}
\caption{Left: Conceptual scheme of transverse atomic momentum variation in $z$-direction in a two-stage SGE (+1/2 component). Right: the transverse momenta in the $p_y-p_z$-plane after magnet 1 and after magnet 2. }
\label{fig:SGEtwostage}
\end{center}
\end{figure}

The transverse momentum vector dependence on the position $x$ inside the SGE is illustrated in Figure \ref{fig:SGEtwostage}: inside and directly behind the magnetic field of magnet 1 all magnetic moments are pointing in the direction of field $F_1$ and have reached the maximum value $\pm p_{z0}$, i.e. the transverse momentum vectors point only in $z$-direction (in Figure \ref{fig:SGEtwostage} the momentum component $p_z$  only for one doublet state is shown, solid line). If magnet 2 is rotated with respect to magnet 1 (angle $\beta$) the atoms at the entrance of magnet 2 are again with $100\%$ probability newly directionally quantized with respect to the field direction $F_2$. Knowing the angle $\beta$ the directional quantization of each atom in magnetic field 1 as well in magnetic field 2 can be unambiguously identified from the location of the momentum vector in the $p_y-p_z$-plane. In the plot ``Momentum after passing magnet 2'' (Figure \ref{fig:SGEtwostage}, right side and down) only the possible final momenta for the $+1/2$ component are shown. For $\beta=90^{\rm o}$ magnet 2 would only contribute to the $p_x$ values. We can conclude that the measurement of the $p_x$ and $p_z$ components yields unambiguous information on the classical path of each atom through a multi-stage SGE. 

Nevertheless, even if a complete alignment of the magnetic moments happens very early in the region where the atoms enter the magnetic field, the precise physical mechanism effecting this alignment remains a mystery to this day, as it had already been to Stern, Einstein, and Ehrenfest.

\section{Directional quantization is an indispensable condition in all quantum physic measurements}

Even after the advent of the new quantum theory of Heisenberg et al. and Schrödinger, 
the results of the SGE remain a mystery in physics (as Feynman had concluded, see above). If Heisenberg's matrix mechanics or the Schrödinger equation would have been known and accepted when the SGE was performed, the outcome of the SGE could have been calculated with these theories and nobody might have been surprised by the results of the SGE! Nevertheless the mechanism of beam splitting remains an unsolved example of the quantum measurement problem. 

Solving the Schrödinger equation for a silver atom in its ground state in a magnetic field one finds directional quantization in perfect agreement with the SGE results. One obtains as solutions (measurable observables) only well-defined angular momentum eigenvalues. These eigenvalues are the angular momentum quantum number $l$ representing an angular momentum component of $l\cdot\hbar$, the so called magnetic quantum number $m$ with an angular momentum component of $m\cdot\hbar$, and the electron spin quantum number component $s$ with an angular momentum of $s\cdot\hbar$. According to the Schrödinger equation, the angular momentum vector can never be determined in quantum physics in direction and length simultaneously. One can only measure the total length of the angular momentum vector together with angular momentum eigenvalues representing projections (i.e. $l$, $m$, and $s$) of the angular momentum vector along the direction of external forces. The components in $x$- and $y$-directions cannot be determined accurately at the same time. 

The new quantum theory does not allow any other projections of the angular momenta for the atoms in a magnetic field, i.e.\ at the point of measurement after the passage of the atoms through the SG apparatus. Thus, the expectation of continuous distributions is a purely classical view. Angular momentum projections smaller than these eigenvalues are forbidden. In quantum physics one can only visualize projections of the angular momentum quantized in $\hbar$ in a direction given by the measurement. This is exactly what Stern and Gerlach observed. The directional quantization is a consequence of the finite size of $\hbar$. Directional quantization must always be present and visible in any measurement data of any quantum system. All structure and dynamics have the signature of angular momentum quantization, and directional quantization is always present in any measurement of a quantum process. 

\begin{figure}
\begin{center}
\includegraphics[scale=0.4]{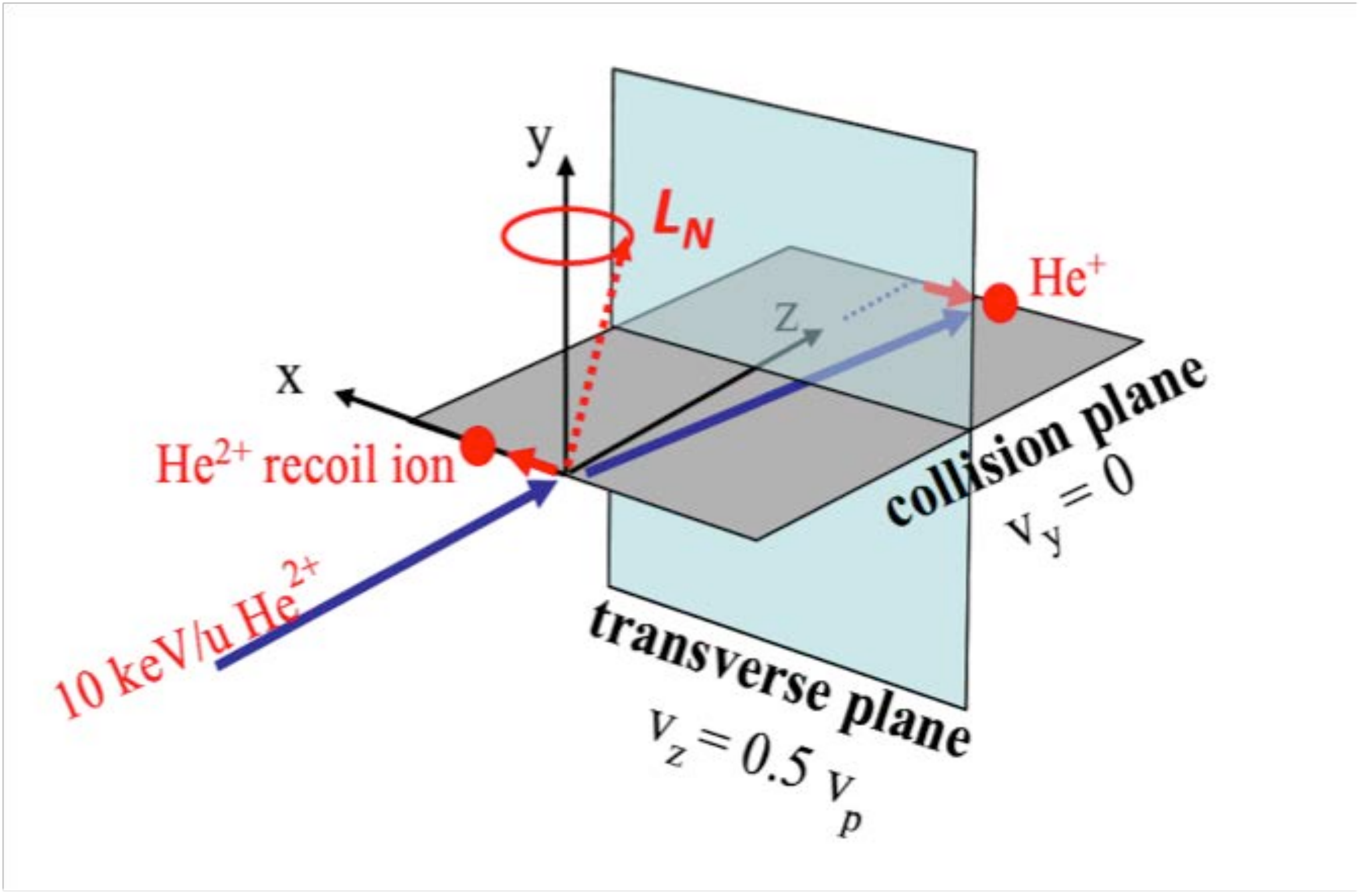}
\caption{Scheme of the measurement \citep{SchmidtLEtal2014Vortices}. Single ions of a well collimated He$^{2+}$ beam are colliding  with single He-Atoms (target) produced in a supersonic jet (very low internal temperature $< 1$K) and it captures one electron from the target atom. The other target electron is ionized into a continuum state. Thus the final scattered projectiles are singly charged and the slow recoil ions (target) are doubly charged. All three particles resulting from the same collision process are detected in coincidence and their momentum is measured with very high resolution ($\delta p<0.02$~a.u.). Thus for each collision the nuclear scattering plane is  determined. $L_{N}$  is the nuclear angular momentum vector. }
\label{fig:HeExp_scheme}
\end{center}
\end{figure}

\begin{figure}
\begin{center}
\includegraphics[scale=0.4]{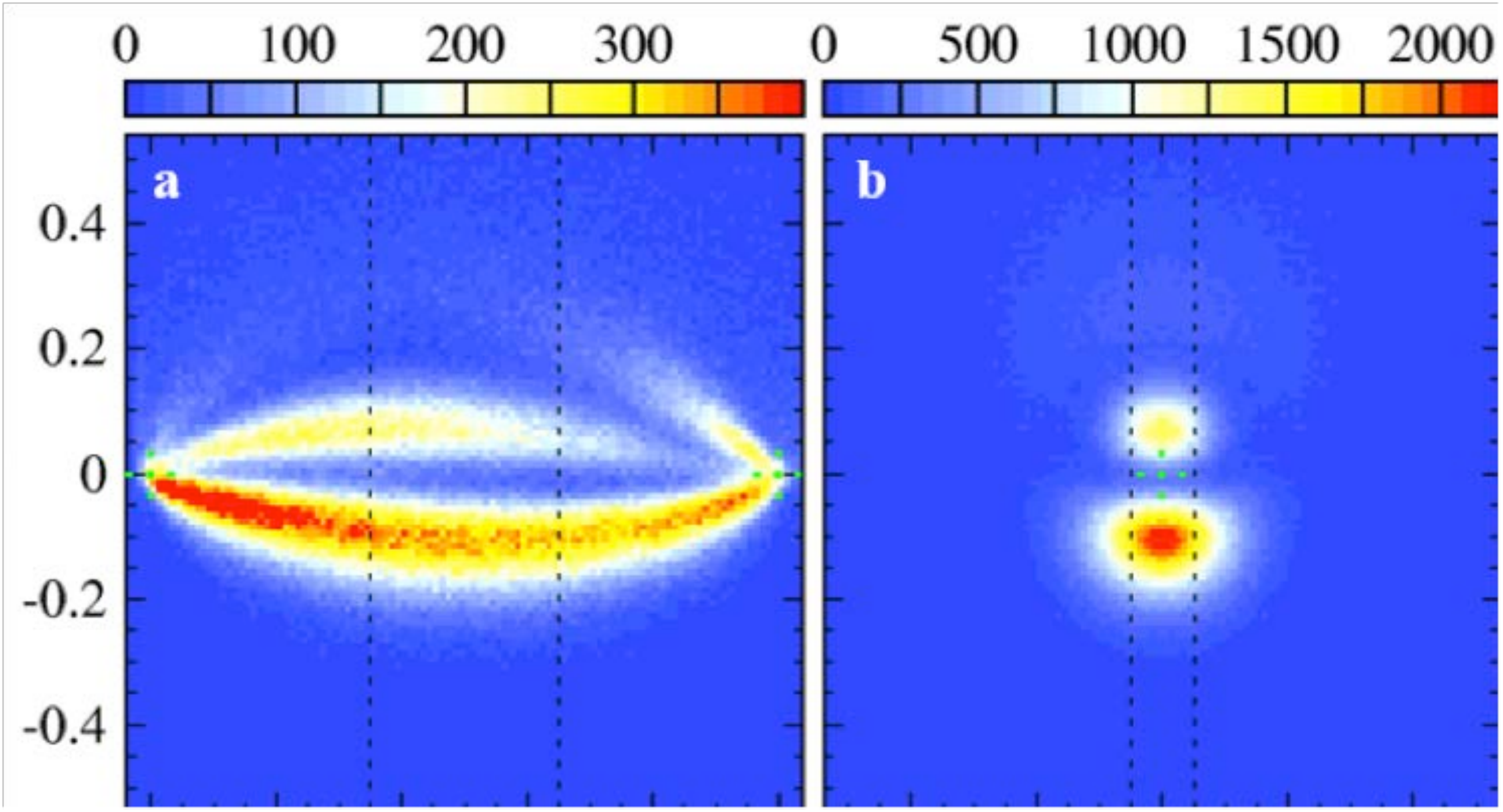}
\caption{Electron momentum distribution at distant collisions: left side in and right side perpendicular to nuclearscattering plane (see text). At distant collisions we observe the well-known two-banana shape \citep{SchmidtLEtal2014Vortices}.}
\label{fig:HeExp_results}
\end{center}
\end{figure}

\begin{figure}
\begin{center}
\includegraphics[scale=0.4]{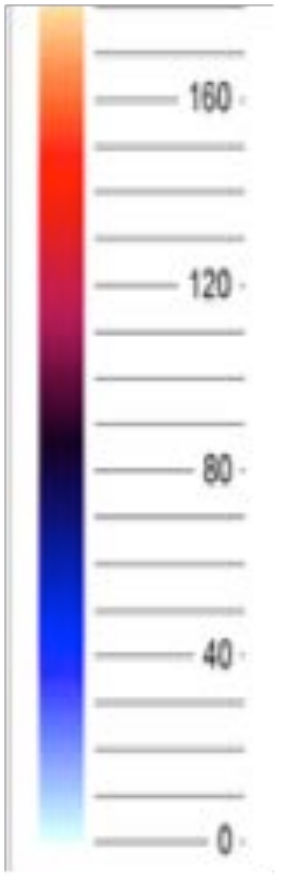}
\includegraphics[scale=0.4]{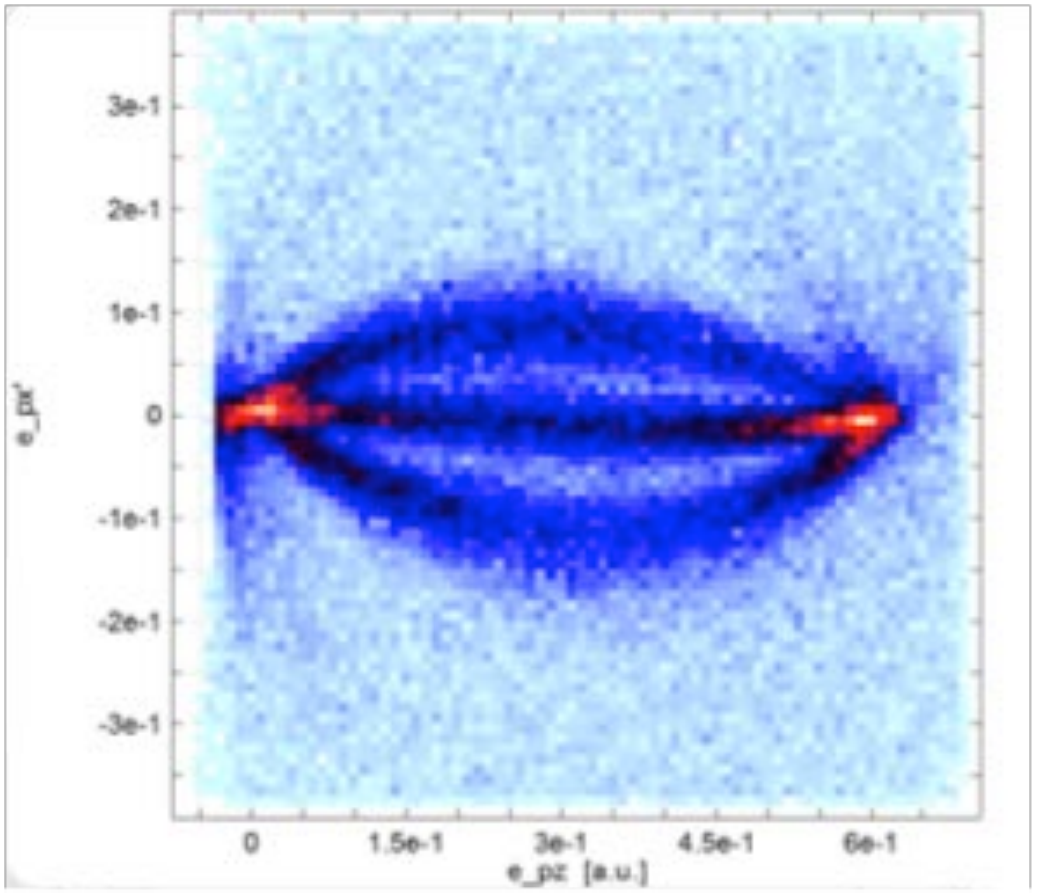}
\includegraphics[scale=0.4]{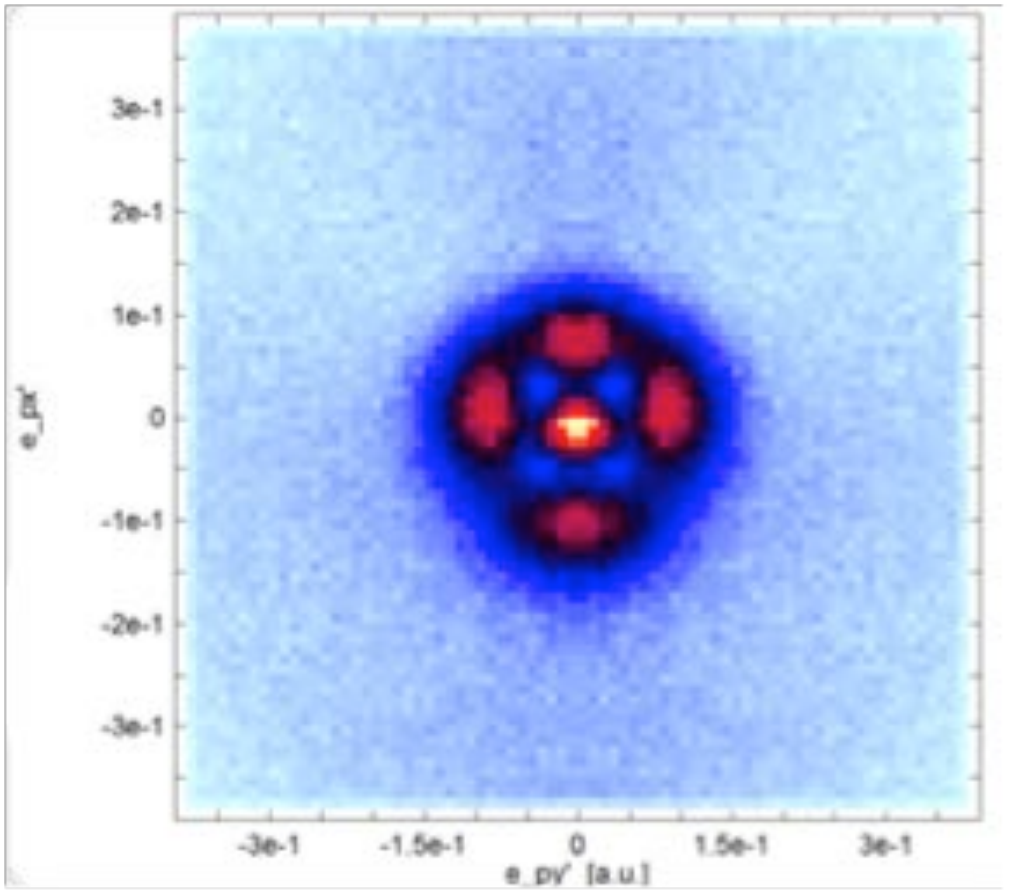}
\includegraphics[scale=0.4]{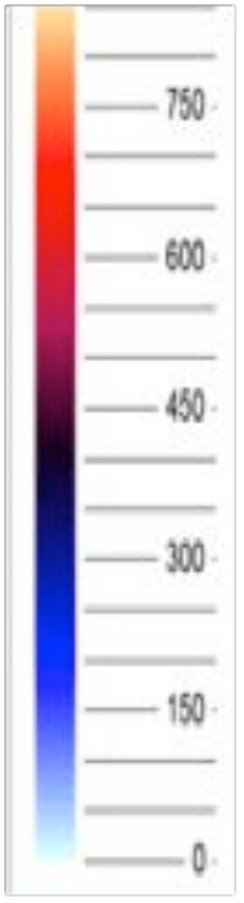}
\caption{The electron emission pattern for close collisons, i.e. larger nuclear scattering angles ($2.25$--$3.25$~mrad). Left side: in nuclear scattering plane, right side: perpendicular to the nuclear scattering plane \citep{SchmidtLEtal2014Vortices}.}
\label{fig:HeExp_results2}
\end{center}
\end{figure}

As an example for directional quantization visible in quantum measurements, we now discuss a recent ion-atom scattering experiment \citep{SchmidtLEtal2014Vortices}. Indeed, the directional quantization effect becomes nicely visible in $10$~keV/u He$^{2+}$+He collisions where the momentum pattern of the emitted electrons with respect to the nuclear scattering plane was measured by recoil-ion He$^{2+}$-electron coincidence. The scheme of the experiment is shown in Figure \ref{fig:HeExp_scheme}. The total nuclear angular momentum vector (approximately $10^{4}\hbar$) points in a direction perpendicular to the nuclear scattering plane. The projections (Figures \ref{fig:HeExp_results}+\ref{fig:HeExp_results2}) show the intensity distributions of the emitted electrons for the regions within in the dotted lines (Figure \ref{fig:HeExp_results}). 

Since electrons from different molecular orbitals can be promoted to the same final momentum state, one final structure in the momentum distribution can be a mixture of two different angular momenta. During the slow collision process the electronic orbits form so-called quasi-molecular states. Electrons can be promoted during the collision process from one quasi-molecular orbital to another one, where also angular momentum is transferred from the nuclear motion to the electronic motion but strictly conserving angular momentum during the collision process from one attosecond to the next one. 

Total angular momentum has no uncertainty with ongoing collision time. The electrons promoted to continuum (free ionized electrons) must be all in well-aligned angular momentum states or mixtures of them and must show directional quantization. In Figure \ref{fig:HeExp_results} the electron emission pattern for distant collisions, i.e. small nuclear deflection angles ($<1.25$~mrad, large impact parameter) are shown. The left hand side presents the electron momentum distribution in the nuclear scattering plane. The momenta are plotted in units of the projectile velocity and the colors reflect the intensities. The right hand side shows the electron momentum distribution structure perpendicular to the scattering plane, which can be explained by the promotion of electrons from the $2p{\pi}_u$ quasi-molecular state to the continuum.
      
Because of the symmetry with respect to the scattering plane there will be no pure $p$ states but only a $p_x$ state which is responsible for electron emission in the scattering plane. The ``banana'' shaped structures can be attributed to the following angular momentum states: The high intensity ``banana'' shaped structures in Figure \ref{fig:HeExp_results} are $\pi$-angular momentum states. The asymmetry originates from a $\sigma$ contribution. In Figure \ref{fig:HeExp_results} (left side) a superposition of $2s\sigma$ and $3d\delta$ contributions is visible where the main quantum numbers are those of the quasi-molecule. The electron promotion proceeds here via a hidden-crossing process and it promotes $1s\sigma$, $3d\sigma$, $5f\sigma$, $7i\sigma$, etc.\ to the continuum.     
      
In Figure \ref{fig:HeExp_results2} the electron emission pattern for close collisions, i.e.\ larger nuclear scattering angles ($2.25$--$3.25$~mrad) are shown; on the left hand side, in nuclear scattering plane, on the right hand side, perpendicular to the nuclear scattering plane. Here the nuclei approach each other in a closer distance (smaller impact parameter) and exchange more momentum. Furthermore the quasi molecular electronic orbitals are more like the united atom orbitals. Other electron promotion processes are allowed and different angular momentum can be exchanged. Even more bananas become visible and the electron emission in the plane perpendicular to the nuclear scattering plane shows significant structures which are only due to quantized angular momentum transfer with directional quantization. 
      
If no electron-recoil coincidence measurement is performed (i.e.\ the outer force direction is randomly distributed), i.e.\ the electron emission is averaged over all non-oriented nuclear collision planes then  no angular momentum induced structure in the electron momentum distribution is visible and directional quantization can not be seen.  As conclusion we can say if the coincidence experiment brings the quantum object into a well defined angular momentum alignment or orientation with respect to the experimental apparatus then directional quantization is always visible and appears as a fundamental feature in any quantum reaction process. 

\section{Almost discovery of the electron spin}

The magnetic properties of atomic matter were a top research topic in physics at the beginning of the 20$^{\rm th}$ century. In the course of disentangling ever more complicated line spectra, the concept of the electronic spin emerged in the mid-twenties \citep{TomonagaS1997Story}, i.e.\ after the original SGE. There were, however, early indications hinting at an electronic spin momentum so that the electron's spin could, perhaps, almost have been found by the time of the original SGE.

To explain the multiplet structures of light emission in the Zeeman effect Sommerfeld concluded in 1920 that there must be another additional hidden magnetic moment in an atom, he called it ``inner quantum number'' (``Innere Quantenzahl'') and named the quantum number of this angular momentum $k$ without, however, specifying its origin \citep{SommerfeldA1920Zahlenmysterium,SommerfeldA1920Gesetze}. 

In addition, in 1921 Arthur H.~Compton (1892--1962) discussed the possibility that the electron itself was magnetic creating additional magnetic moments in an atom \citep{ComptonAH1921electron,StuewerR1975Effect}. From data on atomic magnetism available at the time he concluded that the observed magnetism could not come from groups of atoms such as molecules because this effect was not visible in a displacement of the individual atoms and because Laue diagrams did not show that. Compton concluded that the origin of the magnetism must be inside the atom itself, presumably in the electron itself, which is the ultimate magnetic particle, and he supposed that the electron is spinning like a gyroscope.

Compton considered the behavior of dia- and paramagnetic properties of matter in an outer magnetic field. Applying Lenz's law, he obtained qualitative but not quantitative agreement. Based on Parson's hypothesis \citep{ParsonA1915Theory} of a continuous ring of negative electricity spinning rapidly about an axis perpendicular to its plane, Compton argued that the electron has an isotropic charge distribution with a strong concentration of charge near the center. Analyzing electron track structures in a Wilson chamber, Compton concluded that the electron path shows a spiral tendency, which he interpreted as evidence that the free electron possess magnetic polarity and acts both as a tiny magnet and as an electric charge. This important paper by Compton found very little if any attention in Europe. Thus, the possibility of a magnetic moment arising from a spinning electron was not given any notice in the discussion of the results of the SGE in Europe. 

Today we know that the doublet splitting observed by Stern and Gerlach is due to the electron spin of $\hbar/2$ with a $g$-factor of about $2$. Bohr's authority, however, was obviously so commanding that nobody appears to have had any doubts about Bohr's classical doublet explanation.  It took four more years until it was recognized that the splitting that Gerlach and Stern had seen was a consequence of the electron spin. Already in 1922, Alfred Landé (1888--1976) came very close to the correct interpretation of the doublet splitting in the SGE \citep{LandeA1921Zeemaneffekt2,LandeA1921Zeemaneffekt,LandeA1929Polarisation}. In a paper entitled ``Difficulties of the quantum theory of atomic structure, in particular of a magnetic kind” 
\citep{LandeA1924Schwierigkeiten}, he proposed that the doublet splitting in the SGE would be due to a total angular momentum of silver of $J=1/2$ with projections $m=\pm 1/2$ and a $g$-factor of $2$. However, Landé did not provide any answer nor any speculation as to the origin of the angular momentum of $J=1/2$. He only pointed out that this value of $g$ for spectroscopic $s$ terms is in good agreement with recent magneto-mechanical experiments by \cite{BarnettS1915Magnetization} and \cite{EinsteinAEtAl1915Nachweis}\footnote{For historical literature relevant to the discovery of the gyromagnetic factor of 2 in Einstein-de Haas-type experiments, see \cite[Doc.~215]{CPAE10}.}
 and that Bohr's explanation would be wrong. The possibility of an electron spin was never mentioned in any of Landé's papers before 1925. In addition to Landé's work, the young Heisenberg tried to explain the Zeeman-effect in 1922 by introducing angular momentum values of $1/2$ \citep{HeisenbergW1922Quantentheorie}. But his thesis supervisor Arnold Sommerfeld did not like that idea \cite[secs.~8.4--8.5]{EckertM2013Sommerfled}.
 
Since Landé, Gerlach, and Stern were sitting close to each other in next door offices in the Frankfurt physics building in the years 1918/19--1921, one might surmise that they would had discussed this problem of doublet splitting. None of the three scientists reported such discussions. However, remarks in a letter Stern wrote to Landè on 24.1.1923 one may interpret to the effect that they must have talked about half $\hbar$ quantum numbers. In the letter Stern wrote: 
\begin{quote}
My congratulations for the multiplets and the general Zeeman equation. I am particularly happy that you punch down Heisenberg, I am only afraid that he will save himself by means of some new hypotheses.\footnote{%
``Zu den Multipletts und der allgemeinen Zeemannformel gratuliere ich sehr. Besonders freut mich, daß Sie den Heisenberg absägen; ich befürchte nur, er wird sich einfach mit Hilfe einiger neuer Hypothesen retten. (US-OSF)}
\end{quote}
Clearly, this comment indicates that in early 1923 Stern knew about Landé's work on the Zeeman effect. We may surmise that this knowledge included half $\hbar$ angular momentum ideas and also that he might have connected it to the SGE.

\section{Conclusions}

The SGE performed 1922 by Walther Gerlach and Otto Stern in Frankfurt is a seminal experiment in physics. With our knowledge of today there is no doubt that the SGE was a key benchmark experiment that provided evidence for several of the unexpected basic features of quantum physics. It was performed about 3 years before quantum mechanics was developed which provided a theoretical framework in which the quantum world radically differs from the classical one. The SGE had provided evidence that many of the theoretical hypothesis on quantization in the atomic world were real. 

In this paper we argue that the long-term impact of the SGE is even greater than normally stated in literature. The SGE opened the door for atomic and molecular beam experiments under vacuum conditions exploring the properties of isolated atoms. Before Stern and Gerlach, nobody was able to investigate single atoms. The SGE pioneered and revolutionized atomic and molecular physics. The SG apparatus represents a momentum microscope of the size of a pencil but with a resolution that allowed one to visualize quantum properties inside an atom or a nucleus. It is fully justified that the place in Frankfurt, where the experiment was performed, is now chosen as a ``historic site'' of science in honor of Otto Stern and Walther Gerlach. 

\section*{Acknowledgment}

We acknowledge very helpful discussions with Klaus Blaum, Reinhard Dörner, Michael Eckert, Bretislav Friedrich, Gernot Gruber, Joseph Georg Huber, Bruno Lüthi, Wolfgang Quint, Jan Peter Toennies, and Volkmar Vill. We thank Diana Templeton-Killen for providing private information (Photos, ancestry of Otto Stern etc.) on the Stern Family, Pia Seyler-Dielmann for supporting us with historic literature, and Ruth Speiser-Bär for providing documents and photo material.

\section*{Archival Locations}

\begin{tabbing}
xxxxxxxxxxx\=\kill\\
Gy-DM     \> Deutsches Museum, Archiv, Munich, Germany. \\
Gy-UAF   \> Goethe-Universität Frankfurt, Archiv, Frankfurt, Germany\\
Gy-UBFAZ \> University Library Frankfurt J.C. Senckenberg, Archivzentrum,\\
                \> Frankfurt, Germany\\
Is-AEA      \> Albert Einstein Archives, The Hebrew University of Jerusalem,\\
                 \> Jerusalem, Israel\\
Ne-LeMB    \> Museum Boerhaave, Leyden, The Netherlands \\
Sw-RSAS \> The Royal Swedish Academy of Sciences, Center for History of Science, \\
                \> Stockholm, Sweden \\
Sz-ETHA  \> ETH-Bibliothek Z\"urich, Archives, Zurich, Switzerland \\
US-BL      \>  University of California, Bancroft Library, Berkeley, CA, USA\\
US-NBLA  \> American Institute of Physics,  Niels Bohr Library and Archives, \\
                 \> College Park, MD, USA\\
US-OSF \> Otto Stern papers in possession of the family \\
                  \> (Alan Templeton, Oakland, CA, USA) 
\end{tabbing}

\bibliographystyle{apalike}

\bibliography{SGE_revisited}

\begin{thebibliography}{}

\bibitem[Aaserud and Heilbronn, 2013]{AaserudFEtal2013Love}
Aaserud, F. and Heilbronn, J.~L. (2013).
\newblock {\em {Love, Literature, and the Quantum Atom}}.
\newblock Oxford: Oxford University Press.

\bibitem[Aston, 1919]{AstonF1919Spectrograph}
Aston, F. (1919).
\newblock {A positive ray spectrograph}.
\newblock {\em Philosophical Magazine}, 38:707--714.

\bibitem[Bacciagaluppi and Valentini, 2009]{BacciagaluppiGEtAl2009Crossroads}
Bacciagaluppi, G. and Valentini, A. (2009).
\newblock {\em {Quantum Theory at the Crossroads. Reconsidering the 1927 Solvay
  Conference}}.
\newblock Cambridge: Cambridge University Press.

\bibitem[Barnett, 1915]{BarnettS1915Magnetization}
Barnett, S. (1915).
\newblock {Magnetization by Rotation}.
\newblock {\em Physical Review}, 6:239--270.

\bibitem[Bernstein, 2010]{BernsteinJ2010Experiment}
Bernstein, J. (2010).
\newblock {The Stern-Gerlach Experiment}.
\newblock arXiv:1007.2435.

\bibitem[Bohm, 1993]{BohmA1993Mechanics}
Bohm, A. (1993).
\newblock {\em {Quantum Mechanics. Foundations and Applications}}.
\newblock New York: Springer.

\bibitem[Bohm, 1951]{BohmD1951Quantum}
Bohm, D. (1951).
\newblock {\em {Quantum theory}}.
\newblock Englewood Cliffs, NJ: Prentice-Hall.

\bibitem[Bohr, 1913a]{BohrN1913Constitution1}
Bohr, N. (1913a).
\newblock {On the Constitution of Atoms and Molecules}.
\newblock {\em Philosophical Magazine}, 26(151):1--25.

\bibitem[Bohr, 1913b]{BohrN1913Constitution2}
Bohr, N. (1913b).
\newblock {On the Constitution of Atoms and Molecules. Part II. Systems
  containing only a Single Nucelus}.
\newblock {\em Philosophical Magazine}, 26(153):476--502.

\bibitem[Bohr, 1913c]{BohrN1913Constitution3}
Bohr, N. (1913c).
\newblock {On the Constitution of Atoms and Molecules. Part III. Systems
  containing Several Nuclei}.
\newblock {\em Philosophical Magazine}, 26:857--875.

\bibitem[Bohr, 1949]{BohrN1949Discussion}
Bohr, N. (1949).
\newblock {Discussion with Einstein on Epistemological Problems in Modern
  Physics}.
\newblock In Schilpp, P.~A., editor, {\em Albert Einstein
  Philosopher--Scientist}, pages 199--242. La Salle, Ill.: Open Court.

\bibitem[Born, 1920]{BornM1920Relativitaetstheorie}
Born, M. (1920).
\newblock {\em {Die Relativitätstheorie Einsteins und ihre physikalischen
  Grundlagen}}.
\newblock Berlin (u.a.): Springer.

\bibitem[Compton, 1921]{ComptonAH1921electron}
Compton, A.~H. (1921).
\newblock {The magnetic electron}.
\newblock {\em Franklin Institute Journal}, 192(2):145--155.

\bibitem[Dahms, 2002]{DahmsHJ2002Politics}
Dahms, H.-J. (2002).
\newblock {Appointment Politics and the Rise of Modern Theoretical Physics at
  Göttingen}.
\newblock In Rupke, N., editor, {\em Göttingen and the Development of the
  Natural Sciences}, pages 143--157. Göttingen: Wallstein.

\bibitem[Darrigol, 1992]{DarrigolO1992From}
Darrigol, O. (1992).
\newblock {\em {From "c"-numbers to "q"-numbers: The classical analogy in the
  history of quantum theory}}.
\newblock Berkeley: Univ. of California Press.

\bibitem[Darwin, 1928]{DarwinC1928Motion}
Darwin, C. (1928).
\newblock {Free Motion in the Wave Dynamics}.
\newblock {\em Proceedings of the Royal Society of London A}, 117:258--293.

\bibitem[Debye, 1916]{DebyeP1916Quantenhypothese}
Debye, P. (1916).
\newblock {Quantenhypothese und Zeeman-Effekt}.
\newblock {\em Physikalische Zeitschrift}, 17(20):507--512.

\bibitem[Dempster, 1918]{DempsterAJ1918Method}
Dempster, A. (1918).
\newblock {A New Method of Positive Ray Analysis }.
\newblock {\em Physical Review}, 11(4):316--325.

\bibitem[Devereux, 2015]{DevereuxM2015Reduction}
Devereux, M. (2015).
\newblock {Reduction of the atomic wavefunction in the Stern-Gerlach
  experiment}.
\newblock {\em Canadian Journal of Physics}, 93(11):1382--1390.

\bibitem[Eckert, 2013]{EckertM2013Sommerfled}
Eckert, M. (2013).
\newblock {\em {Arnold Sommerfeld. Atomphysiker und Kulturbote 1868--1951. Eine
  Biografie}}.
\newblock Göttingen: Wallstein.

\bibitem[Einstein, 1905]{Einstein1905i}
Einstein, A. (1905).
\newblock {Über einen die Erzeugung und Verwandlung des Lichtes betreffenden
  heuristischen Gesichtspunkt}.
\newblock {\em Annalen der Physik}, 17:132--148.
\newblock Reprinted in \cite[Doc.~14, pp.~150--169]{CPAE02}.

\bibitem[Einstein and Ehrenfest, 1922]{EinsteinAEtAl1922Bemerkungen}
Einstein, A. and Ehrenfest, P. (1922).
\newblock {Quantentheoretische Bemerkungen zum Experiment von Stern und
  Gerlach}.
\newblock {\em Zeitschrift für Physik}, 11:31--34.
\newblock Reprinted in \cite[Doc.~315]{CPAE13}.

\bibitem[Einstein and Haas, 1915]{EinsteinAEtAl1915Nachweis}
Einstein, A. and Haas, W. J.~d. (1915).
\newblock {Experimenteller Nachweis der Ampèreschen Molekularströme}.
\newblock {\em Deutsche Physikalische Gesellschaft. Verhandlungen},
  17(8):152--170.

\bibitem[Estermann and Stern, 1933a]{EstermannIEtal1933Moment}
Estermann, I. and Stern, O. (1933a).
\newblock {Magnetic moment of the deuton}.
\newblock {\em Nature}, 133:911.

\bibitem[Estermann and Stern, 1933b]{EstermannIEtal1933Ablenkung}
Estermann, I. and Stern, O. (1933b).
\newblock {Über die magnetische Ablenkung von isotopen Wasserstoffmolekülen
  und das magnetische Moment des ``Deutons''}.
\newblock {\em Zeitschrift für Physik}, 86:132--134.

\bibitem[Estermann and Stern, 1933c]{EstermannIEtal1933Ablenkung2}
Estermann, I. and Stern, O. (1933c).
\newblock {Über die magnetische Ablenkung von Wasserstoffmolekülen und das
  magnetische Moment des Protons II.}
\newblock {\em Zeitschrift für Physik}, 85:17--24.

\bibitem[Feynman, 1963]{FeynmanR1963Lectures}
Feynman, R. (1963).
\newblock {\em {The Feynman Lectures on Physics}}.
\newblock : Addison-Wesley.

\bibitem[Franca, 2009]{FrancaH2009Phenomenon}
Franca, H.~M. (2009).
\newblock {The Stern-Gerlach Phenomenon According to Classical
  Electrodynamics}.
\newblock {\em Foundations of Physics}, 39:1177--1190.

\bibitem[French and Taylor, 1978]{FrenchAPEtal1978Introduction}
French, A. and Taylor, E.~F. (1978).
\newblock {\em {An Introduction to Quantum Physics}}.
\newblock Boca Raton, FL: CRC Press.

\bibitem[Fricke, nd]{FrickeHNoDateJahre}
Fricke, H. (n.d.).
\newblock {\em {150 Jahre Physikalischer Verein Frankfurt a.M.}}
\newblock Ljubljana, Yugoslavia: CGB Delo.

\bibitem[Friedrich and Herschbach, 1998]{FriedrichBEtal1998Star}
Friedrich, B. and Herschbach, D. (1998).
\newblock {Otto Stern's Lucky Star}.
\newblock {\em Daedalus}, 127(1):165--191.

\bibitem[Friedrich and Herschbach, 2003]{FriedrichBEtal2003Stern}
Friedrich, B. and Herschbach, D. (2003).
\newblock {Stern and Gerlach: How a Bad Cigar Helped Reorient Atomic Physics}.
\newblock {\em Physics Today}, 56(12):53--59.

\bibitem[Friedrich and Herschbach, 2005]{FriedrichBEtal2005Stern}
Friedrich, B. and Herschbach, D. (2005).
\newblock {Stern and Gerlach at Frankfurt: Experimental Proof of Space
  Quantization}.
\newblock In Trageser, W., editor, {\em Stern-Stunden. Höhepunkte Frankfurter
  Physik}. Frankfurt: University of Frankfurt, Fachbereich Physik.

\bibitem[Frisch and Segrè, 1933]{FrischOEtal1933Einstellung}
Frisch, O. and Segrè, E. (1933).
\newblock {Über die Einstellung der Richtungsquantelung. II.}
\newblock {\em Zeitschrift für Physik}, 80:610--616.

\bibitem[Frisch and Stern, 1933a]{FrischOEtal1933Ablenkunga}
Frisch, O. and Stern, O. (1933a).
\newblock {Über die magnetische Ablenkung von Wasserstoffmolekülen und das
  magnetische Moment des Protons}.
\newblock {\em Leipziger Vorträge}, 6:36--42.

\bibitem[Frisch and Stern, 1933b]{FrischOEtal1933Ablenkung}
Frisch, O. and Stern, O. (1933b).
\newblock {Über die magnetische Ablenkung von Wasserstoffmolekülen und das
  magnetische Moment des Protons I.}
\newblock {\em Zeitschrift für Physik}, 85:4--16.

\bibitem[Fü{\ss}l, 1998]{FuesslW1998Nachlass}
Fü{\ss}l, W., editor (1998).
\newblock {\em {Der wissenschaftliche Nachla{\ss} von Walther Gerlach}}.
\newblock München: Deutsches Museum.

\bibitem[Gerlach, 1925]{GerlachW1925Richtungsquantelung}
Gerlach, W. (1925).
\newblock {Über die Richtungsquantelung im Magnetfeld II}.
\newblock {\em Annalen der Physik}, 76:163--197.

\bibitem[Gerlach, 1969a]{GerlachW1969Stern}
Gerlach, W. (1969a).
\newblock {Otto Stern zum Gedenken}.
\newblock {\em Physikalische Blätter}, 25(9):412--413.

\bibitem[Gerlach, 1969b]{GerlachW1969Entdeckung}
Gerlach, W. (1969b).
\newblock {Zur Entdeckung des ``Stern-Gerlach-Effektes''}.
\newblock {\em Physikalische Blätter}, 25(10):472.

\bibitem[Gerlach and Stern, 1921]{GerlachWEtal1921Nachweis}
Gerlach, W. and Stern, O. (1921).
\newblock {Der experimentelle Nachweis des magnetischen Moments des
  Silberatoms}.
\newblock {\em Zeitschrift für Physik}, 8:110--111.

\bibitem[Gerlach and Stern, 1922a]{GerlachWEtal1922Moment}
Gerlach, W. and Stern, O. (1922a).
\newblock {Das magnetische Moment des Silberatoms}.
\newblock {\em Zeitschrift für Physik}, 9:353--355.

\bibitem[Gerlach and Stern, 1922b]{GerlachWEtal1922Nachweis}
Gerlach, W. and Stern, O. (1922b).
\newblock {Der experimentelle Nachweis der Richtungsquantelung im Magnetfeld}.
\newblock {\em Zeitschrift für Physik}, 9:349--352.

\bibitem[Gerlach and Stern, 1924]{GerlachWEtal1924Richtungsquantelung}
Gerlach, W. and Stern, O. (1924).
\newblock {Über die Richtungsquantelung im Magnetfeld}.
\newblock {\em Annalen der Physik}, 74:673--699.

\bibitem[Gomis and Pérez, 2016]{GomisP2016Effects}
Gomis, P. and Pérez, A. (2016).
\newblock {Decoherence effects in the Stern-Gerlach experiment using matrix
  Wigner functions}.
\newblock {\em Physical Review A}, 94:012103.

\bibitem[Gordon et~al., 1955]{GordonJEtal1955Maser}
Gordon, J., Zeiger, H., and Townes, C. (1955).
\newblock {The Maser---New Type of Microwave Amplifier, Frequency Standard, and
  Spectrometer}.
\newblock {\em Physical Review}, 99(4):1264--xxx.

\bibitem[Güttinger, 1932]{GuettingerP1932Verhalten}
Güttinger, P. (1932).
\newblock {Das Verhalten von Atomen im magnetischen Drehfeld}.
\newblock {\em Zeitschrift für Physik}, 73:169--184.

\bibitem[Heinrich and Bachmann, 1989]{HeinrichREtal1989Gerlach}
Heinrich, R. and Bachmann, H.-R., editors (1989).
\newblock {\em {Walther Gerlach. Physiker-Lehrer-Organisator}}.
\newblock München: Deutsches Museum.

\bibitem[Heisenberg, 1922]{HeisenbergW1922Quantentheorie}
Heisenberg, W. (1922).
\newblock {Zur Quantentheorie der Linienstruktur und der anomalen
  Zeemaneffekte}.
\newblock {\em Zeitschrift für Physik}, 8:273 -- 297.

\bibitem[Heisenberg, 1927]{HeisenbergW1927Inhalt}
Heisenberg, W. (1927).
\newblock {Über den anschaulichen Inhalt der quantentheoretischen Kinematik
  und Mechanik}.
\newblock {\em Zeitschrift für Physik}, 43(3):172--198.

\bibitem[Hermansphan et~al., 2000]{HermansphanNEtal2000Observation}
Hermansphan, N., Haffner, H., Kluge, H.-J., Quint, W., Stahl, S., Verdú, J.,
  and Werth, G. (2000).
\newblock {Observation of the continuous Stern-Gerlach effect on an electron
  bound in an atomic ion}.
\newblock {\em Physical Review Letters}, 84:427--.

\bibitem[Huber, 2014]{HuberJG2014Gerlach}
Huber, J.~G. (2014).
\newblock {\em {Walther Gerlach (1888--1979) und sein Weg zum erfolgreichen
  Experimentalphysiker bis etwa 1925}}.
\newblock {Phd-dissertation}, {LMU München}.
\newblock dated 6 August 2014.

\bibitem[Inguscio, 2006]{InguscioM2006Comment}
Inguscio, M. (2006).
\newblock {Comment on the scientific paper no. 6: `Oriented atoms in a variable
  magnetic field'}.
\newblock In Bassani, G., editor, {\em Ettore Majorna. Scientific Papers},
  pages 133--136. Bologna, Berlin: Società Italian die Fisica and Springer.

\bibitem[Kellogg et~al., 1939]{KelloggMEtal1939Moments}
Kellogg, J., Rabi, I., Ramsey, N., and Zacharias, J. (1939).
\newblock {The Magnetic Moments of the Proton and the Deuteron. The
  Radiofrequency Spectrum of $H_2$ in Various Magnetic Fields}.
\newblock {\em Physical Review}, 56(8):728--743.

\bibitem[Kormos~Buchwald et~al., 2012]{CPAE13}
Kormos~Buchwald, D., Illy, J., Rosenkranz, Z., and Sauer, T., editors (2012).
\newblock {\em {The Collected Papers of Albert Einstein. Vol.13. The Berlin
  Years: Writings \& Correspondence, January 1922--March 1923}}.
\newblock Princeton: Princeton University Press.

\bibitem[Kormos~Buchwald et~al., 2015]{CPAE14}
Kormos~Buchwald, D., Illy, J., Rosenkranz, Z., Sauer, T., and Moses, O.,
  editors (2015).
\newblock {\em {The Collected Papers of Albert Einstein. Vol.14. The Berlin
  Years: Writings \& Correspondence, April 1923--May 1925}}.
\newblock Princeton: Princeton University Press.

\bibitem[Kormos~Buchwald et~al., 2009]{CPAE12}
Kormos~Buchwald, D., Rosenkranz, Z., Sauer, T., Illy, J., Holmes, V.~I., and
  and, editors (2009).
\newblock {\em {The Collected Papers of Albert Einstein. Vol.12. The Berlin
  Years: Correspondence, January--December 1921}}.
\newblock Princeton: Princeton University Press.

\bibitem[Kormos~Buchwald et~al., 2006]{CPAE10}
Kormos~Buchwald, D., Sauer, T., Rosenkranz, Z., Illy, J., and Holmes, V.~I.,
  editors (2006).
\newblock {\em {The Collected Papers of Albert Einstein. Vol.10. The Berlin
  Years: Correspondence, May--December 1920; and Supplementary Correspondence,
  1909--1920}}.
\newblock Princeton: Princeton University Press.

\bibitem[Kragh, 2012]{KraghH2012Bohr}
Kragh, H. (2012).
\newblock {\em {Niels Bohr and the Quantum Atom. The Bohr Model of Atomic
  Structure 1913--1925}}.
\newblock Oxford: Oyford University Press.

\bibitem[Kuhn, 1978]{KuhnT1978Theory}
Kuhn, T.~S. (1978).
\newblock {\em {Black-body theory and the quantum discontinuity, 1894-1912}}.
\newblock New York: Oxford University Press.

\bibitem[Landé, 1921a]{LandeA1921Zeemaneffekt2}
Landé, A. (1921a).
\newblock {Über den anomalen Zeemaneffekt (II. Teil)}.
\newblock {\em Zeitschrift für Physik}, 7(1):398--405.

\bibitem[Landé, 1921b]{LandeA1921Zeemaneffekt}
Landé, A. (1921b).
\newblock {Über den anomalen Zeemaneffekt (Teil I)}.
\newblock {\em Zeitschrift für Physik}, 5(4):231--241.

\bibitem[Landé, 1924]{LandeA1924Schwierigkeiten}
Landé, A. (1924).
\newblock {Schwierigkeiten in der Quantentheore des Atombaues, besonders
  magnetischer Art}.
\newblock {\em Physikalische Zeitschrift}, 24(1):442--xxx.

\bibitem[Landé, 1929]{LandeA1929Polarisation}
Landé, A. (1929).
\newblock {Polarisation von Materiewellen}.
\newblock {\em Die Naturwissenschaften}, 17(32):634--637.

\bibitem[Langmuir, 1925]{LangmuirI1925Effects}
Langmuir, I. (1925).
\newblock {Thermioinc Effects Caused by Vapours of Alkali Metals}.
\newblock {\em Proceedings of the Royal Society of London A}, 107:61--79.

\bibitem[Mackintosh, 1983]{MackintoshAR1983Experiment}
Mackintosh, A.~R. (1983).
\newblock {The Stern-Gerlach experiment, electron spin and intermediate quantum
  mechanics}.
\newblock {\em European Journal of Physics}, 4:97--106.

\bibitem[Majorana, 1932]{MajoranaE1932Atomi}
Majorana, E. (1932).
\newblock {Atomi orientati in campo magnetico variablie}.
\newblock {\em Nuevo Cimento}, 9:43--50.
\newblock No.6 in ``Scientific Papers'', ed. Bassani, 2007.

\bibitem[Mehra and Rechenberg, 1982]{MehraJEtAl1982Development1}
Mehra, J. and Rechenberg, editors (1982).
\newblock {\em {The Historical Development of Quantum Theory. Vol. 1 in 2
  parts. The Quantum Theory of Planck, Einstein and Sommerfeld: Its Foundation
  and the Rise of Its Difficulties 1900-1925}}.
\newblock New York, Heidelberg, Berlin: Springer Verlag.

\bibitem[Nida-Rümelin, 1982]{NidaRuemelinM1982Bibliographie}
Nida-Rümelin, M., editor (1982).
\newblock {\em {Bibliographie Walther Gerlach. Veröffentlichungen von
  1912--1979}}.
\newblock München: Deutsches Museum.

\bibitem[Parson, 1915]{ParsonA1915Theory}
Parson, A. (1915).
\newblock {A Magneton Theory of the Structure of the Atom}.
\newblock {\em Smithsonian Miscellaneous Collections}, 65(11):1--80.
\newblock iisued November 29, 1915, volume is dated 1916.

\bibitem[Pauli, 1979]{PauliW1979Briefwechsel}
Pauli, W. (1979).
\newblock {\em {Wissenschaftlicher Briefwechsel mit Bohr, Einstein Heisenberg
  u.a. Bannd 1: 1919--1929}}.
\newblock New York: Springer.

\bibitem[Phipps and Stern, 1932]{PhippsTEtal1932Einstellung}
Phipps, T.~E. and Stern, O. (1932).
\newblock {Über die Einstellung der Richtungsquantelung}.
\newblock {\em Zeitschrift für Physik}, 73:185--191.

\bibitem[Pié~i Valls, 2015]{Pie2015Experiment}
Pié~i Valls, B. (2015).
\newblock {\em {L'experiment d'Stern i Gerlach en el seu context teòric: la
  història d'una reorientació}}.
\newblock {Ph.D. thesis}, {Universitat de Barcelona}.
\newblock English summary and conclusions in ch. 7, pp. 209--222.

\bibitem[Planck, 1899]{PlanckM1899Strahlungsvorgaenge}
Planck, M. (1899).
\newblock {Über irreversible Strahlungsvorgänge and (Fünfte Mittheilung /
  Schluss)}.
\newblock {\em Königlich Preu{\ss}ische Akademie der Wissenschaften.
  Sitzungsberichte}, 25:440--480.

\bibitem[Platt, 1990]{PlattD1992Analysis}
Platt, D.~E. (1990).
\newblock {A modern analysis of the Stern-Gerlach experiment}.
\newblock {\em American Journal of Physics}, 60(4):306--308.

\bibitem[Popper, 1982]{PopperK1982Theory}
Popper, K. (1982).
\newblock {\em {Quantum theory and the schism in physics}}.
\newblock Totowa, NJ: Rowman and Littlefield.
\newblock from the ``Postscript to the Logic of Scientific Discovery'', pp.
  22--23.

\bibitem[Popper, 1989]{PopperK1989Logic}
Popper, K. (1989).
\newblock {\em {The Logic of Scientifi Discovery}}.
\newblock : Hutchinson.

\bibitem[Rabi, 1929]{RabiI1929Methode}
Rabi, I. (1929).
\newblock {Zur Methode der Ablenkung von Molekülstrahlen}.
\newblock {\em Zeitschrift für Physik}, 54:190--197.

\bibitem[Rabi et~al., 1934]{RabiIEtal1934Moment}
Rabi, I., Kellogg, J., and Zacharias, J. (1934).
\newblock {The Magnetic Moment of the Proton}.
\newblock {\em Physical Review}, 46(3):157--163.

\bibitem[Rabi et~al., 1939]{RabiIEtal1939Method}
Rabi, I., Millman, S., Kusch, P., and Zacharias, J. (1939).
\newblock {The Molecular Beam Resonance Method for Measuring Nuclear Magnetic
  Moments. The Magnetic Moments of 3Li6, 3Li7 and 9F19}.
\newblock {\em Physical Review}, 55(6):526--535.

\bibitem[Reinisch, 1999]{ReinischG1999Experiment}
Reinisch, G. (1999).
\newblock {Stern-Gerlach experiment as the pioneer---and probably the
  simplest---quantum entanglement test?}
\newblock {\em Physics Letters A}, 259:427--430.

\bibitem[Ribeiro, 2010]{RibeiroJ2010Was}
Ribeiro, J. E.~A. (2010).
\newblock {Was the Stern-Gerlach Phenomenon Classically Described?}
\newblock {\em Foundations of Physics}, 40:1779--1782.

\bibitem[Sackur, 1911]{SackurO1912Anwendung}
Sackur, O. (1911).
\newblock {Die Anwendung der kinetischen Theorie der Gase auf chemische
  Probleme}.
\newblock {\em Annalen der Physik}, 36:958--980.

\bibitem[Sackur, 1913]{SackurO1913Bedeutung}
Sackur, O. (1913).
\newblock {Die universelle Bedeutung des sog. elementaren Wirkungsquantums}.
\newblock {\em Annalen der Physik}, 40:67--86.

\bibitem[Sauer, 2016]{SauerT2016Perspectives}
Sauer, T. (2016).
\newblock {Multiple Perspectives on the Stern-Gerlach Experiment}.
\newblock In Sauer, T. and Scholl, R., editors, {\em The Philosophy of
  Historical Case Studies}, pages 251--263. : Springer.

\bibitem[Schmidt et~al., 2014]{SchmidtLEtal2014Vortices}
Schmidt, L., Goihl, C., Metz, D., Schmidt-Böcking, H., Dörner, R.,
  Ovchinnikov, S., Macek, J., and Schultz, D. (2014).
\newblock {Vortices Associated with the Wave Function of a Single Electron
  Emitted in Slow Ion-Atom Collisions}.
\newblock {\em Physical Review Letters}, 112:083201--xxx.

\bibitem[Schmidt-Böcking and Reich, 2011]{SchmidtBoeckingHEtal2011Stern}
Schmidt-Böcking, H. and Reich, K. (2011).
\newblock {\em {Otto Stern. Physiker, Querdenker, Nobelpreisträger}}.
\newblock Frankfurt/Main: Societäts-Verlag.

\bibitem[Schwinger, 2001]{SchwingerJ2001Mechanics}
Schwinger, J. (2001).
\newblock {\em {Quantum mechanics: symbolism of atomic measurements. Ed. by
  Bertold-Georg Englert}}.
\newblock Berlin: Springer.

\bibitem[Schütz, 1969]{SchuetzW1969Erinnerungen}
Schütz, W. (1969).
\newblock {Persönliche Erinnerungen an die Entdeckung des
  Stern-Gerlach-Effektes}.
\newblock {\em Physikalische Blätter}, 25(8):343--345.

\bibitem[Scully et~al., 1987]{ScullyMOEtal1987Theory}
Scully, M.~O., Lamb, Willis~E., J., and Barut, A. (1987).
\newblock {On the Theory of the Stern-Gerlach Apparatus}.
\newblock {\em Foundations of Physics}, 17(6):575--583.

\bibitem[Scully et~al., 1978]{ScullyMOEtal1978Reduction}
Scully, M.~O., Shea, R., and McCullen, J. (1978).
\newblock {State reduction in quantum mechanics: a calculational example}.
\newblock {\em Physics Reports}, 43(13):485--498.

\bibitem[Segrè, 1993]{SegreE1993Mind}
Segrè, E. (1993).
\newblock {\em {A mind always in motion. The autobiography of Emilio Segrè}}.
\newblock Berkeley: University of California Press.

\bibitem[Sommerfeld, 1916]{SommerfeldA1916Theorie}
Sommerfeld, A. (1916).
\newblock {Zur Theorie des Zeeman-Effekts der Wasserstofflinien, mit einem
  Anhang über den Stark-Effek and }.
\newblock {\em Physikalische Zeitschrift}, 17(20):491--507.

\bibitem[Sommerfeld, 1920a]{SommerfeldA1920Gesetze}
Sommerfeld, A. (1920a).
\newblock {Allgemeine spektroskopische Gesetze, insbesondere ein
  magnetooptischer Zerlegungssatz}.
\newblock {\em Annalen der Physik}, 63:221--263.

\bibitem[Sommerfeld, 1920b]{SommerfeldA1920Zahlenmysterium}
Sommerfeld, A. (1920b).
\newblock {Ein Zahlenmysterium in der Theorie des Zeeman-Effektes}.
\newblock {\em Die Naturwissenschaften}, 8(4):61--64.

\bibitem[Sommerfeld, 1921]{SommerfeldA1921Atombau2}
Sommerfeld, A. (1921).
\newblock {\em {Atombau und Spektrallinien}}.
\newblock Braunschweig: Vieweg.

\bibitem[Sommerfeld, 1924]{SommerfeldA1924Atombau4}
Sommerfeld, A. (1924).
\newblock {\em {Atombau und Spektrallinien}}.
\newblock Braunschweig: Vieweg.

\bibitem[Stachel, 1989]{CPAE02}
Stachel, J., editor (1989).
\newblock {\em {The Collected Papers of Albert Einstein. Vol. 2. The Swiss
  Years: Writings, 1900--1909}}.
\newblock Princeton, N.J.: Princeton University Press.

\bibitem[Stern, 1920a]{SternO1920Messung}
Stern, O. (1920a).
\newblock {Eine direkte Messung der thermischen Molekulargeschwindigkeit}.
\newblock {\em Zeitschrift für Physik}, 2:49--56.

\bibitem[Stern, 1920b]{SternO1920Messung2}
Stern, O. (1920b).
\newblock {Eine direkte Messung der thermischen Molekulargeschwindigkeit}.
\newblock {\em Physikalische Zeitschrift}, 21:582.

\bibitem[Stern, 1920c]{SternO1920Nachtrag}
Stern, O. (1920c).
\newblock {Nachtrag zu meiner Arbeit: ``Eine direkte Messung der thermischen
  Molekulargeschwindigkeit''}.
\newblock {\em Zeitschrift für Physik}, 3:417--421.

\bibitem[Stern, 1921]{SternO1921Weg}
Stern, O. (1921).
\newblock {Ein Weg zur experimentellen Prüfung der Richtungsquantelung}.
\newblock {\em Zeitschrift für Physik}, 7:249--253.

\bibitem[Stuewer, 1975]{StuewerR1975Effect}
Stuewer, R.~H. (1975).
\newblock {\em {The Compton effect : turning point in physics}}.
\newblock New York: Science History Publications.

\bibitem[Tetrode, 1912a]{TetrodeH1912Konstante1}
Tetrode, H. (1912a).
\newblock {Die chemische Konstante der Gase und das elementare
  Wirkungsquantum}.
\newblock {\em Annalen der Physik}, 38:434--442.

\bibitem[Tetrode, 1912b]{TetrodeH1912Konstante2}
Tetrode, H. (1912b).
\newblock {Die chemische Konstante der Gase und das elementare Wirkungsquantum
  II.}
\newblock {\em Annalen der Physik}, 39:255--256.

\bibitem[Toennies et~al., 2011]{ToenniesJEtal2011Stern}
Toennies, J.~P., Schmidt-Böcking, H., Friedrich, B., and Lower, J.~C. (2011).
\newblock {Otto Stern (1888--1969): The founding father of experimental atomic
  physics}.
\newblock {\em Annalen der Physik}, 523(12):1045--1070.

\bibitem[Tomonaga, 1997]{TomonagaS1997Story}
Tomonaga, S.-i. (1997).
\newblock {\em {The Story of Spin}}.
\newblock Chicago: The University of Chicago Press.

\bibitem[Trageser, 2011]{TrageserW2011Effekt}
Trageser, W. (2011).
\newblock {\em {Der Stern-Gerlach-Effekt. Genese, Entwicklung und
  Rekonstruktion eines Grundexperimentes der Quantentheorie 1916--1926}}.
\newblock {Ph.D. Thesis}, {Johann Wolfgang Goethe-Universität Frankfurt}.

\bibitem[Unna and Sauer, 2013]{UnnaIEtal2013Einstein}
Unna, I. and Sauer, T. (2013).
\newblock {Einstein, Ehrenfest, and the quantum measurement problem}.
\newblock {\em Annalen der Physik}, 525(1-2):A15--A19.

\bibitem[Weinert, 1995]{WeinertF1995Theory}
Weinert, F. (1995).
\newblock {Wrong Theory - Right Experiment: The Significance of the
  Stern-Gerlach Experiments}.
\newblock {\em Studies in the History and Philosophy of Modern Physics},
  26(1):75--86.
\newblock [Ztschr. im MPIWG vorh.].

\bibitem[Wennerström and Westlund, 2012]{WennerstroemHEtal2012experiment}
Wennerström, H. and Westlund, P. (2012).
\newblock {The Stern-Gerlach experiment and the effects of spin relaxation}.
\newblock {\em Phys. Chem. Chem. Phys.}, 14:1677--1684.

\bibitem[Wennerström and Westlund, 2013]{WennerstroemHEtal2013measurements}
Wennerström, H. and Westlund, P. (2013).
\newblock {On Stern-Gerlach coincidence measruements and their applications to
  Bell's theorem}.
\newblock {\em Physics Essays}, 26:174--180.

\bibitem[Wennerström and Westlund, 2014]{WennerstroemHEtal2014Interpretation}
Wennerström, H. and Westlund, P. (2014).
\newblock {Interpretation versus explanation in the description of the
  Stern-Gerlach experiment}.
\newblock preprint.

\bibitem[Zeeman, 1896]{ZeemanP1896Invloed}
Zeeman, P. (1896).
\newblock {Over den Invloed eener Magnetisatie op den Aard van het door een
  Stof uitgezonden Licht}.
\newblock {\em Koninklijke Akademie van Wetenschappen te Amsterdam. Section of
  Sciences. Proceedings}.

\bibitem[Zeeman, 1897]{ZeemanP1897Influence}
Zeeman, P. (1897).
\newblock {On the Influence of Magnetism on the Nature of the Light Emitted by
  a Substance}.
\newblock {\em Philosophical Magazine}, 43:226--239.

\end{thebibliography}

\end{document}